\documentclass[a4paper,11pt]{article}
\usepackage[a4paper, total={6in, 8in}]{geometry}
\usepackage{amsmath}
\usepackage{amssymb}
\usepackage{latexsym}
\usepackage{empheq}
\usepackage{subcaption}
\usepackage{authblk}
\usepackage[numbers]{natbib}
\usepackage{graphicx}
\usepackage{url}
\usepackage{xcolor}
\definecolor{newcolor}{rgb}{.8,.349,.1}

\title{A fully Eulerian solver for the simulation of multiphase flows with solid bodies: 
application to surface gravity waves}

\author[1]{Francesco De Vita}
\author[1,2]{Filippo De Lillo}
\author[3,4]{Roberto Verzicco}
\author[1,2]{Miguel Onorato}
\affil[1]{Department of  Physics, University of Turin, via Pietro Giuria 1, 10125 Turin, Italy}
\affil[2]{INFN, via Pietro Giuria 1, 10125 Turin, Italy}
\affil[3]{Department of Industrial Engineering, University of Rome Tor Vergata, via del Politecnico 1, Rome, Italy}
\affil[4]{PoF grup, University of Twente, The Netherlands}

\begin{document}
\maketitle
\begin{abstract}
In this paper a fully Eulerian solver for the study of multiphase flows for simulating the 
propagation of surface gravity waves over submerged bodies is presented. 
We solve the incompressible Navier--Stokes equations coupled with the 
volume of fluid technique for the modeling 
of the liquid phases with the inteface, an immersed body method for the solid bodies and an 
iterative strong--coupling procedure for the fluid--structure interaction. 
The flow incompressibility is enforced via the solution of a Poisson equation which, owing to
the density jump across the interfaces of the liquid phases, has to resort to 
the splitting procedure of \citeauthor{Dodd2014} \cite{Dodd2014}. 

The solver is validated through comparisons against classical test cases for fluid--structure 
interaction like migration of particles in pressure--driven channel, multiphase flows, `water exit'
of a cylinder and a good agreement is found for all tests. Furthermore, we show the application 
of the solver to the case of a surface gravity wave propagating over a submerged reversed pendulum 
and verify that the solver can reproduce the energy exchange between the wave and the pendulum. 
Finally the three--dimensional spilling breaking of a wave induced by a submerged sphere is considered.
\end{abstract}

\section{Introduction}

Multiphase flows with fluid--structure interaction (FSI) are found in many physical 
and engineering areas. The computational modeling of these systems poses several challenges 
and different methods have been developed in the past to solve the two problems, separately. 
Multiphase flows involve two immiscible fluids separated by an interface; an accurate description 
of the motion of this surface is necessary to correctly predict the overall flow dynamics;
examples of these flows are droplets and bubbles but also ocean waves. 

On the other hand, in FSI problems the motion of a body determines the flow, via the boundary conditions,
and, in turn, the hydrodynamic loads cause the motion of the body. 
Both classes of problems imply a time--dependent configuration and possibly changes of flow topology;
in order to avoid the generation of body--fitted meshes, a common approach has been used which
requires the solution on a Cartesian grid of the Naviers---Stokes equations coupled with 
additional models to describe the presence of the interfaces and solid bodies.

Immersed boundary methods (IBM) \cite{peskin1972} are an efficient solution for 
flows around moving bodies since the governing equations are solved on a fixed Cartesian grid and 
an additional forcing term in the momentum equation imposes the no--slip boundary conditions at the
immersed surface. The direct forcing approach \cite{Fadlun2000} is particularly attractive due to the 
small limitation on the timestep and computational overhead although it suffers from spurious 
pressure oscillations \cite{Lee2011}, which can be alleviated by mass corrections \cite{Kim2001} 
or extrapolation of the fluid field inside the solid \cite{Yang2006}. Alternatively, 
an Eulerian/Lagrangian approach can be used \cite{uhlmann2005} by describing the immersed surface
through a network of Lagrangian markers and transfering the forcing between the Lagrangian and Eulerian 
grid by a discrete delta function; recently this approach has been modified introducing a 
moving--least--square interpolation for the spreading of the forcing \cite{Vanella2014}. A drawback 
of this method is that, differently from in \cite{Fadlun2000}, it requires an iteration of the forcing 
step since the force spreading modifies the Eulerian points associated to each Lagrangian marker. 
The direct forcing approach of \cite{Fadlun2000} has been effectively used for simulations of flows
around complex rigid bodies \cite{deTullio2009,DeVita2016,Viola2020} because in this case the rigid 
motion depends only on the net hydrodynamic force and the oscillations smooth out in the surface 
integration. On the contrary, for deformable bodies it is necessary to use a Lagrangian approach 
to avoid local large displacements induced by pointwise force fluctuations \cite{DeTullio2016}.

Regarding multiphase flows, one of the most used approache is the one--fluid formulation. 
Within this method the two fluid phases are considered as a single contiuum with variable material 
properties which are discontinuous across the interface; a source term is then introduced in the 
momentum equation to take into account the presence of surface tension (see for example Chapter 2 
of \cite{tryggvason2011direct}). An additional advection equation is however required to track 
in time the motion of the interface. The front--tracking method, first developed by \cite{unverdi1992}, 
is an Eulerian/Lagrangian method used to solve the Navier--Stokes equations on a Cartesian grid and 
it employs a moving mesh for the interface separating the fluids; since the additional grid deforms, 
remeshing is necessary during the calculation. To avoid this computationally expensive step, 
an alternative approach is the front--capturing method, which is fully Eulerian and handles topological 
changes without the need of additional grids. This yields a more flexible numerical method that allows 
for a more efficient parallelization with respect to the Eulerian/Lagrangian counterpart. 
The Eulerian methods are basically the volume--of--fluid (VoF) \cite{scardovelli1999} and the level--set 
(LS) methods \cite{sethian1996}. VoF identifies the different phases introducing a color function;
 its main advantage is the intrinsic mass conservation but the geometric representation of the interface 
(the normal vector and the curvature) is not trivial. LS, instead, identifies the fluids using a 
signed distance function which allows an efficient and accurate evaluation of normal vector and 
curvature but it suffers from mass loss. 

A common approach to advance in time the incompressible Navier--Stokes equations is the 
fractional--step method \cite{Kim1985} which consist of two main steps: the first computes a provisional
velocity field by the momentum equation with the pressure at the old timestep and then projects
the velocity onto a solenoidal field to enforce mass conservation. This correction step requires 
the solution of a Poisson equation for the pressure with the density as coefficient. Unfortunately, 
in all the methods previously described for multiphase flows the density is a function of the interface
position, hence the Poisson equation has space--dependent coefficients which makes impossible the
use of fast direct solvers (FDS) \cite{swarztrauber1977}. This issue is usually coped by the use of 
multigrid methods \cite{Chen1999,Popinet2009,Yang2009} that, however, are computationally expensive and
whose convergence rate is problem dependent. Recently, \citeauthor{Dodd2014} \cite{Dodd2014} proposed 
a new procedure to approximate the variable coefficient Poisson equation by a constant coefficient 
counterpart: this is accomplished by introducing an approximation of the unknown new pressure 
at timestep $n+1$ which allows to split the updated pressure and density in the Poisson equation. 
They have shown that, by computing an approximated pressure as an extrapolation from the previous 
timestep, the overall $2^{nd}$ order accuracy of the method is preserved. Nevertheless, the combination 
of this splitting technique with the IBM is non trivial owing to the presence of the pressure in the 
right hand side of the Poisson equations. \citeauthor{Frantzis2019} \cite{Frantzis2019} have proposed, 
for a stationary solid, a local reconstruction of the pressure gradient at the boundary in order 
to avoid the use of solid nodes in the Poisson equations and to yield a decoupled solution of the 
fluid and solid domains.

To the best of the authors' knowledge there are no solvers in literature which simulate the 
Navier--Stokes equations for multiphase flows with moving solid boundaries employing FDS for 
the solution of the Poisson equation. In order to fill this gap, the aim of this work is to present 
an efficient fully Eulerian solver for the simulations of multiphase flows with moving solid bodies. 
To achieve this goal we couple several ingredients: the presence of solid bodies in a fluid phase
is described by the direct forcing approach \cite{Fadlun2000} combined by an interpolation scheme 
similar to that proposed by \citeauthor{Balaras2004} \cite{Balaras2004}; the solid phase dynamics
is integrated by a $4^{th}$ order predictor--corrector scheme, based on \citeauthor{Hamming1959} 
\cite{Hamming1959} method. For the multiphase flow, we use the Multi-dimensional Tangent of Hyperbola 
Interface Capturing (MTHINC) VoF developed by \cite{Ii2012}. Although the method can be used for generic 
geometries by the ray--tracing procedure described in \cite{Iacca}, here the analysis is limited to 
solid bodies which can be described by analytical formulas (\emph{e.g.} spheres, ellipsoids \emph{etc}):
the proposed method is optimal for simulations of suspensions in multiphase flows or surface gravity 
waves with submerged bodies.

The paper is structured as follows: in Section \ref{sec:formulation} we introduce the mathematical 
framework for the physical system and in Section \ref{sec:numerics} we describe all the details
of the numerical solver. In section \ref{sec:results} we present some validation tests and two 
applications for a surface gravity wave propagating over a submerged reversed pendulum in two and 
three dimensions.

\section{Formulation}\label{sec:formulation}
\begin{figure}
  \centering
  \includegraphics[width=0.5\textwidth]{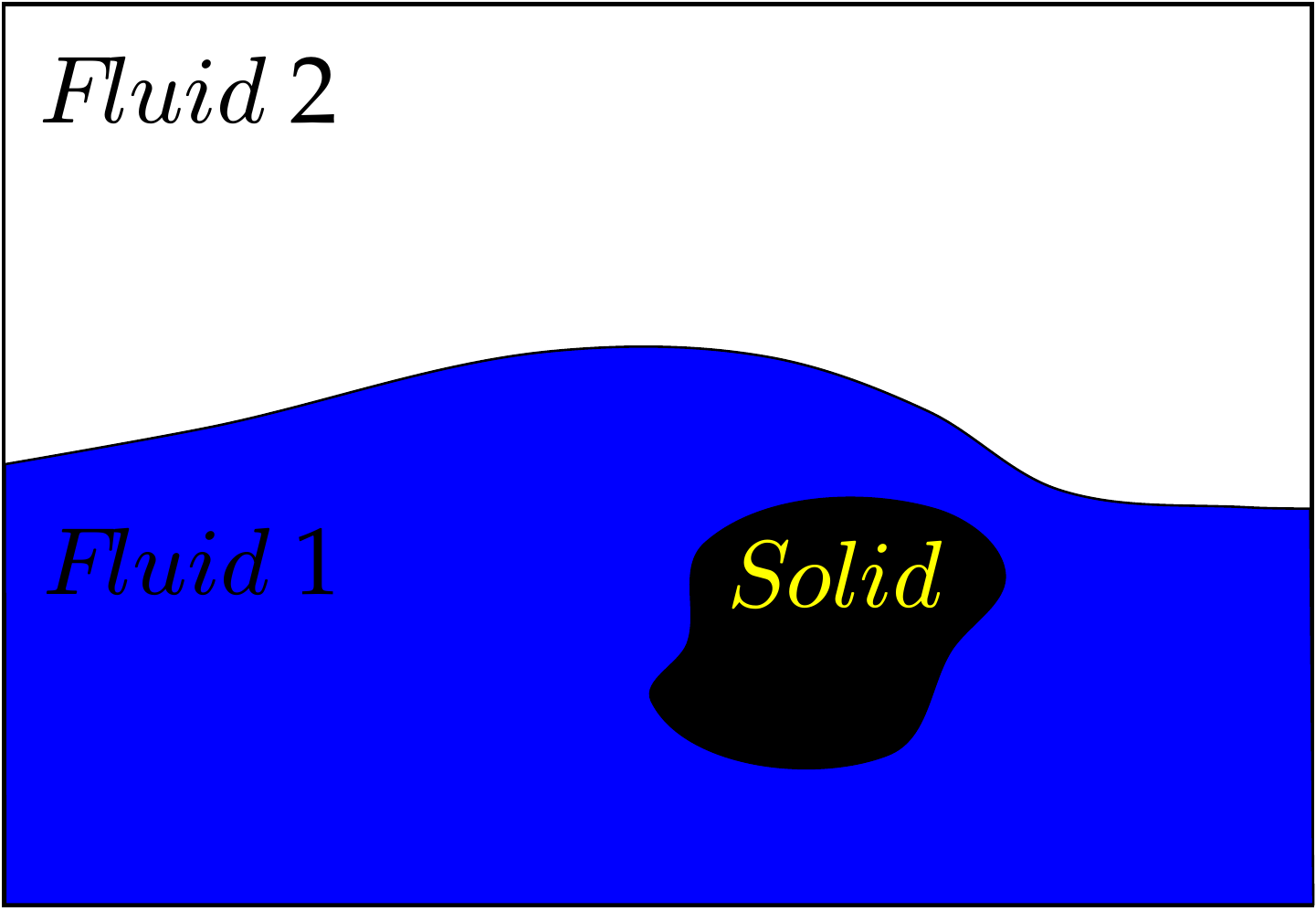}
  \caption{Sketch of a three--phase flow: a surface gravity wave 
           propagation with a submerged body.\label{fig:sketch}}
\end{figure}
The basic features of the problem are sketched in figure \ref{fig:sketch} 
with two immiscible fluids separated by an interface and one or more 
solid bodies.
We use the one--fluid formulation of the Navier--Stokes equations 
and introduce an indicator (or color) 
function $\mathcal{H}$ to identify the fluids such that $\mathcal{H}$ is 
$1$ in one fluid and $0$ in the other; the interface is tracked by an 
additional advection equation for the indicator function. 
To account for the solid phase, we use the IBM which adds a forcing term 
to the momentum equation to impose the no--slip boundary condition at the 
fluid/solid boundary. The full system of equations reads:

\begin{subequations}
  \label{eqn:fullsystem}
  \begin{empheq}[left=\empheqlbrace]{align}
    &\nabla \cdot \vec{u} = 0 \label{eqn:mass}\\
    &\rho\left(\frac{\partial \vec{u}}{\partial t} + \vec{u}\cdot \nabla \vec{u}\right) = -\nabla p + \nabla \cdot \mathbf{\tau} + \rho\vec{g} + \rho\vec{f} + \vec{f}_{\sigma} \label{eqn:momentum}\\
    &\frac{\partial \mathcal{H}}{\partial t} + \vec{u} \cdot \nabla \mathcal{H} = 0 \label{eqn:advection}\\
    &m_n\frac{d \vec{V_n}}{dt} = \vec{F_n} \label{eqn:veltra}\\
    &\frac{d\vec{X_n}}{dt} = \vec{V_n}\label{eqn:postra}\\
    &\mathbf{I}_n\frac{d \vec{\Omega}_n}{dt} = \vec{M}_n \label{eqn:velrot}\\
    &\frac{d \vec{\Theta}_n }{dt} = \vec{\Omega}_n \label{eqn:posrot}\
  \end{empheq}
\end{subequations}
where $\vec{u}$ is the velocity vector, $p$ the pressure, $\rho$ and 
$\mu$ the density and dynamic viscosity of the fluid, $\mathbf{\tau}$ 
the viscous stress tensor that for a Newtonian fluid in indicial form
is $\tau_{ij} = 2\mu D_{ij} = \mu(\partial u_i/\partial x_j + 
\partial u_j/\partial x_i)$. $\vec{g}$ is the gravity, $\vec{f}$ the forcing 
term which enforces the no--slip boundary condition at the immersed boundaries 
(described in section \ref{sec:sbt}) and $\mathcal{H}$ the indicator function. The 
force $\vec{f}_{\sigma}$ is the surface tension force acting at the interface between 
the two fluids and directed normal to the interface. The vectors $\vec{X}_n$ and $\vec{\Theta}_n$ 
are the linear and angular positions
of the center of mass of the $n^{th}$ solid body, while $\vec{V}_n$ and 
$\vec{\Omega}_n$ its linear and angular velocity, $\vec{F}_n$ and $\vec{M}_n$ 
the external force and moment acting on the body. $m_n$ is the mass and 
$\mathbf{I}_n$ the moment of inertia tensor; the subscript $n$ spans all
the solid bodies ($1 \leq n \leq N$). 
Note that lower case letters ($u$, $p$, \emph{etc.}) are used
for Eulerian fluid variables while upper case letters ($V$, $X$, \emph{etc.}) 
for Lagrangian solid variables.

The fluid properties are constant within each phase and therefore are determined
only by the function $\mathcal{H}$, as specified below. 

The system \eqref{eqn:fullsystem} has $5$ unknowns for the fluid phases (three
velocity components, the pressure and the interface function $\mathcal{H}$) and 
$12N$ unknowns for the solid phase ($6$ positions and $6$ velocities). 
The momentum equation \eqref{eqn:momentum} depends on the interface location via 
the material properties $\rho$ and $\mu$ and on the location of the solid phase 
via the forcing term $\vec{f}$. 
The solid body equations \eqref{eqn:veltra}-\eqref{eqn:posrot}, instead, depend 
on the flow field via the loads $\vec{F}_n$ and $\vec{M}_n$:
\begin{subequations}
  \label{eqn:loads}
  \begin{align}
	  &\vec{F}_n = \int_{S_n} \left(\mathbf{\tau}\cdot\vec{n} - p\vec{n}\right)\,dS \\
	  &\vec{M}_n = \int_{S_n} \left[\left(\mathbf{\tau}\cdot\vec{n} - p\vec{n}\right)\times \vec{r}\right]\,dS
  \end{align}
\end{subequations}

where $S_n$ is the surface of the $n^{th}$ solid body, $\vec{n}$ the local outward
normal and $\vec{r}$ the local distance from its center of mass. This makes the 
equations of the system \eqref{eqn:fullsystem} strongly coupled and some 
instabilities can arise when integrating them numerically \cite{Borazjani2008}. 
In the next section we describe the steps to solve the full system of equations.

\section{Numerical solver}\label{sec:numerics}
\subsection{Flow solver}
We advance in time the momentum equation \eqref{eqn:momentum} by
a fractional--step method \cite{Chorin1967}: first a provisional
non--soleinodal velocity field $\vec{u}^*$ is computed using an explicit 
$2^{nd}$ order 
Adams--Bashfort scheme for the non--linear and viscous terms, yielding the 
semi--discrete equation for the timestep $l$:
\begin{equation}
  \label{eqn:projection}
  \frac{\vec{u}^*-\vec{u}^l}{\Delta t} = -\frac{1}{\rho^{l+1}}\nabla p^l 
+ \frac{3}{2}\vec{H}^l - \frac{1}{2}\vec{H}^{l-1} + \vec{g}^l + \vec{f}^l,
\end{equation}
where the term $\vec{H}^l$ is given by
\begin{equation}
  \label{eqn:rhs}
  \vec{H}^l = -\vec{u}^l\cdot \nabla \vec{u}^l + 
\frac{1}{\rho^{l+1}}\nabla \cdot \left(2\mu^{l+1}\mathbf{D}^l\right)
\end{equation}
with $\mathcal{D}^l$ the deformation rate tensor at timestep $l$ 
$D_{ij}^l = (\partial u_i^l/\partial x_j + \partial u_j^l/\partial x_i)$ 
and the material properties $\rho^{l+1}$ and $\mu^{l+1}$ are obtained after 
the advection of the interface, as described in the next section.
The correct velocity field at timestep $l+1$ is obtained by computing a 
scalar function $\phi$ such that
\begin{equation}
  \label{eqn:correction}
  \vec{u}^{l+1} = \vec{u}^* - \frac{\Delta t}{\rho^{l+1}}\nabla \phi^l;
\end{equation}
by summing equation \eqref{eqn:projection} and \eqref{eqn:correction} it 
can be seen that the new pressure is given by
\begin{equation}
  \label{eqn:newp}
  p^{l+1} = p^{l} + \phi^l,
\end{equation}
hence $\phi^l$ represents the pressure increment at the timestep $l$, 
that can be computed taking the divergence of equation \eqref{eqn:correction}:
\begin{equation}
  \label{eqn:variablepoisson}
  \nabla \cdot \left( \frac{\Delta t}{\rho^{l+1}}\nabla \phi^l \right) = 
\nabla \cdot \vec{u}^*.
\end{equation}
This is a variable coefficient Poisson equation because of the presence 
of the density $\rho^{l+1}$ inside the divergence operator. In order to use 
FDS, following \citeauthor{Dodd2014}\cite{Dodd2014}, the pressure gradient in 
the momentum equation is expressed as
\begin{equation}
  \label{eqn:splitting}
  \frac{\nabla p^{l+1}}{\rho^{l+1}} \rightarrow \frac{\nabla p^{l+1}}{\rho_0} 
+ \left(\frac{1}{\rho^{l+1}}-\frac{1}{\rho_0}\right)\nabla \hat{p} = 
  \frac{\nabla p^{l}}{\rho_0} + \frac{\nabla \phi^l}{\rho_0} + 
\left(\frac{1}{\rho^{l+1}}-\frac{1}{\rho_0}\right)\nabla \hat{p},
\end{equation}
where $\hat{p}$ is an approximation of the pressure $p^{l+1}$ and 
$\rho_0 = \min\left(\rho_1,\rho_2\right)$ is a constant density. 
Using this surrogate for the pressure gradient the momentum equation 
yields the following relation for the provisional velocity field
\begin{equation}
  \label{eqn:ustar}
  \frac{\vec{u}^*-\vec{u}^l}{\Delta t} = -\frac{1}{\rho_0}\nabla p^l - \left(\frac{1}{\rho^{l+1}}-\frac{1}{\rho_0}\right)\nabla \hat{p} + \frac{3}{2}\vec{H}^l - \frac{1}{2}\vec{H}^{l-1} + \vec{g}^l + \vec{f}^l,
\end{equation}
with the new Poisson equation for the pressure increment
\begin{equation}
  \label{eqn:poisson}
  \nabla^2 \phi^l = \frac{\rho_0}{\Delta t}\nabla \cdot \vec{u}^*
\end{equation}
and the new velocity correction
\begin{equation}
  \label{eqn:newcorrection}
  \vec{u}^{l+1} = \vec{u}^* - \frac{\Delta t}{\rho_0}\nabla \phi^l.
\end{equation}
The splitting procedure \eqref{eqn:splitting} decouples the pressure gradient 
from the variable density and allows the use of FDS schemes for the solution of 
the Poisson equation. 
\citeauthor{Frantzis2019} \cite{Frantzis2019} proposed a local reconstruction 
of the pressure gradient at the immersed boundaries to avoid the use of points 
inside the solid regions. However, with the formulation 
\eqref{eqn:ustar}--\eqref{eqn:newcorrection}, this is not 
effective since, while the reconstruction proposed in \cite{Frantzis2019} 
provides a solenoidal velocity field only in the fluid domain, here the solid 
phase is moving and information from the solid region would enter anyway 
the fluid domain because of the dynamics. On the other hand we will see that,
because of the fluid/structure interaction, at least two steps (predictor and
corrector) and often some iterations are necessary to advance one time interval
$\Delta t$, therefore the errors at the immersed boundaries are largely 
reduced by the multi--step procedure.

It is worth noticing that in the phase with lower density there is no approximation 
(the last term on the RHS of \eqref{eqn:splitting} cancels out) whereas in the 
phase with the higher density the quality of the approximation depends on how 
close $\hat{p}$ is to $p^{l+1}$: in the limit $\hat{p} \equiv p^{l+1}$ again 
there is no approximation. \citeauthor{Dodd2014}\cite{Dodd2014} have shown that 
extrapolating $\hat{p}$ from the steps $l$ and $l-1$, that for a constant 
time step integration corresponds to $\hat{p} = 2p^l - p^{l-1}$, the approximation 
\eqref{eqn:splitting} preserves the second--order time accuracy of the scheme. 
Equations \eqref{eqn:ustar}, \eqref{eqn:poisson} and \eqref{eqn:newcorrection} 
are the steps needed to advance the flow field from timestep $l$ to $l+1$. 
All the spatial derivatives are approximated by second--order accurate 
finite--difference scheme except for the viscous terms of equation \eqref{eqn:rhs} 
where the fifth--order WENO is used \cite{Shu2009}; this is necessary to avoid 
oscillations since viscosity jumps are localized across the interface over three 
grid nodes. Due to the explicit treatment of the non--linear and viscous terms, 
the timestep is restricted for stability reasons
\begin{equation}
  \Delta t = \min\left(CFL\frac{\Delta}{|u_i|_{\max}}, C_{\nu}\frac{\Delta ^2}{\nu}\right),
\end{equation}
where $\Delta$ is the grid size, $CFL$ the Courant--Friedrichs--Lewy parameter
and $C_{\nu}$ a coefficient depending on the number of spatial dimensions
($1/4$ in 2D and $1/6$ in 3D). Note that even though the theoretical $CFL$ 
condition for the Adams--Bashfort scheme is $CFL \leq 1$, in practice the 
maximum is $CFL \approx 0.3$ \cite{VanderPoel2015}. As it will be shown in 
the section of the results, this condition can be further restricted for large 
density ratios bacause of the approximation \eqref{eqn:splitting}.

\subsection{Interface tracking}

The method employed for the interface tracking is that proposed in 
\cite{Rosti2018b}, based on the MTHINC originally developed by \cite{Ii2012}. 
Here we briefly describe only the key features of the method and the reader 
is referred to \cite{Rosti2018b} for a detailed description of the scheme
and the relevant validation tests. 
The cell averaged value of the indicator 
function $\mathcal{H}$ is defined as the volume fraction of a fluid phase (or
volume of fluid, VoF) within a control volume $\delta V$
\begin{equation}
  \mathcal{F}\left(\vec{x},t\right) \equiv \frac{1}{\delta V} 
\int_{\delta V}\mathcal{H}\left(\vec{x},t\right) dV,
\end{equation}
being $0 \le \mathcal{F}\left(\vec{x},t\right) \le 1$. 
The advection equation for the VoF function is then
\begin{equation}
  \label{eqn:vof}
  \frac{\partial \mathcal{F}}{\partial t} + 
\nabla \cdot \left(\vec{u}\mathcal{H}\right) = \mathcal{F}\vec{u}.
\end{equation}
The key point of the MTHINC method is the approximation of the color 
function by an hyperbolic tangent
\begin{equation}
  \mathcal{H}(\vec{X}) \approx \hat{\mathcal{H}}(\vec{X}) = 
\frac{1}{2}(1 + \text{tanh}(\beta(P(\vec{X})+q))),
\end{equation}
where $\vec{X} \in [0,1]$ is a local coordinate system, $\beta$ a sharpness 
parameter, $q$ a shift and $P$ a three--dimensional surface function which can 
be either linear (plane) or quadratic (curved surface) at no additional cost. 
This discretization allows to solve the fluxes of equation \eqref{eqn:vof} by 
integration of the approximated color function in each computational cell. 
The material properties of the two fluids are connected to the VoF function 
$\mathcal{F}$ as follows:
\begin{equation}
  \label{eqn:properties}
  \begin{aligned}
    \rho(\vec{x},t) = \rho_1\mathcal{F}(\vec{x},t) + \rho_0(1-\mathcal{F}(\vec{x},t)), \\
    \mu(\vec{x},t) = \mu_1\mathcal{F}(\vec{x},t) + \mu_0(1-\mathcal{F}(\vec{x},t)),
  \end{aligned}
\end{equation}
where the subscript $1$ stands for phase where $\mathcal{F}$ is equal to $1$ and 
the subscript $0$ for the other phase. The surface tension force $\vec{f}_{\sigma}$ can
be numercially described usgin the Continuum Surface Force method \cite{Brackbill1992} for which
$\vec{f}_{\sigma} = \sigma \kappa \nabla \mathcal{H}$, with $\sigma$ the surface tension coefficient
and $\kappa$ the local curvature of the interface. The solver has been tested in 
\cite{Rosti2018b} and recently used in \cite{Rostihst2019,devita2019}.

\subsection{Solid boundary treatment}\label{sec:sbt}
\begin{figure}
  \centering
  \begin{subfigure}{0.4\textwidth}
    \includegraphics[width=\textwidth]{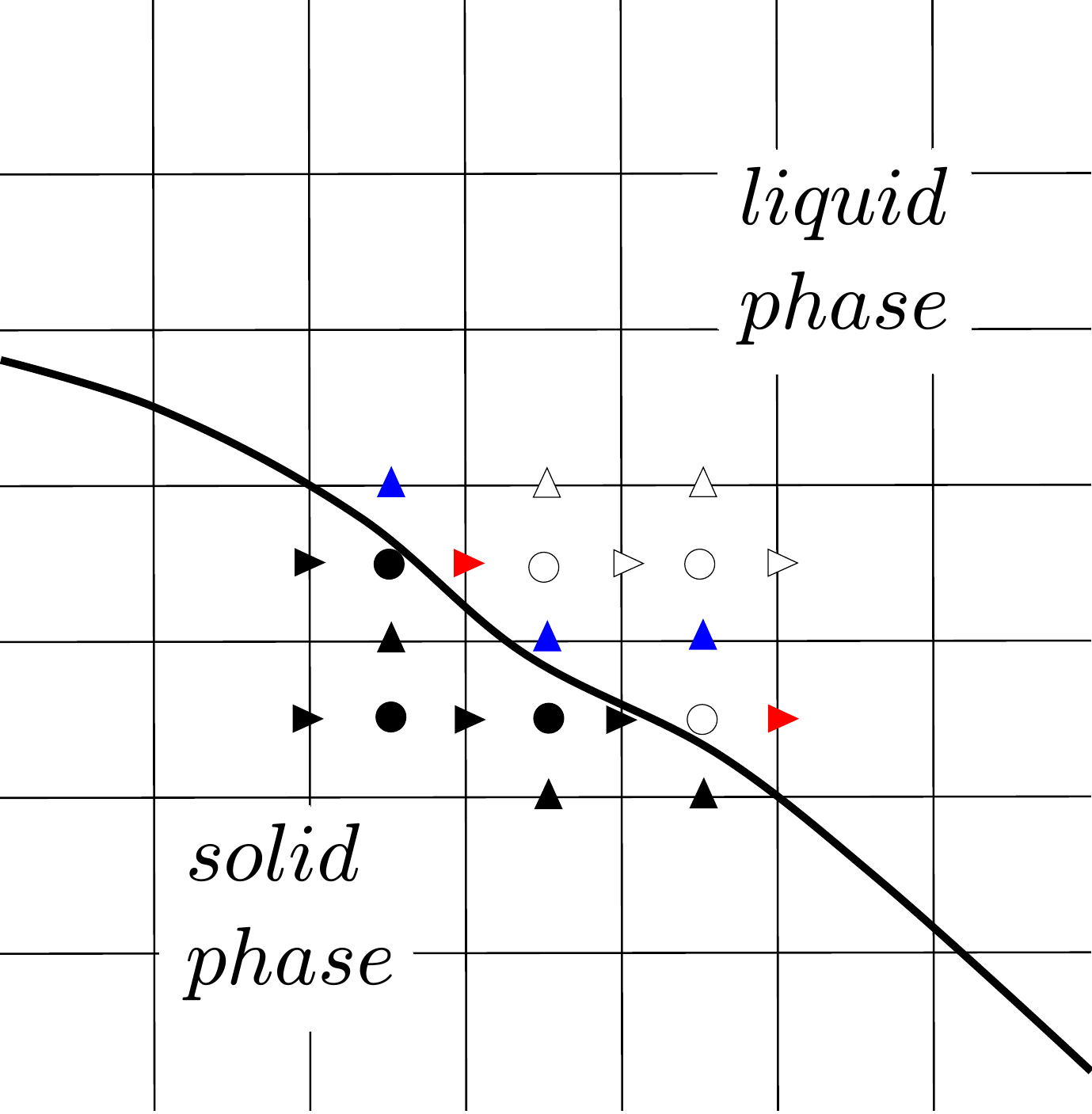}
    \caption{\label{fig:forcing}}
  \end{subfigure}
  \begin{subfigure}{0.4\textwidth}
    \includegraphics[width=\textwidth]{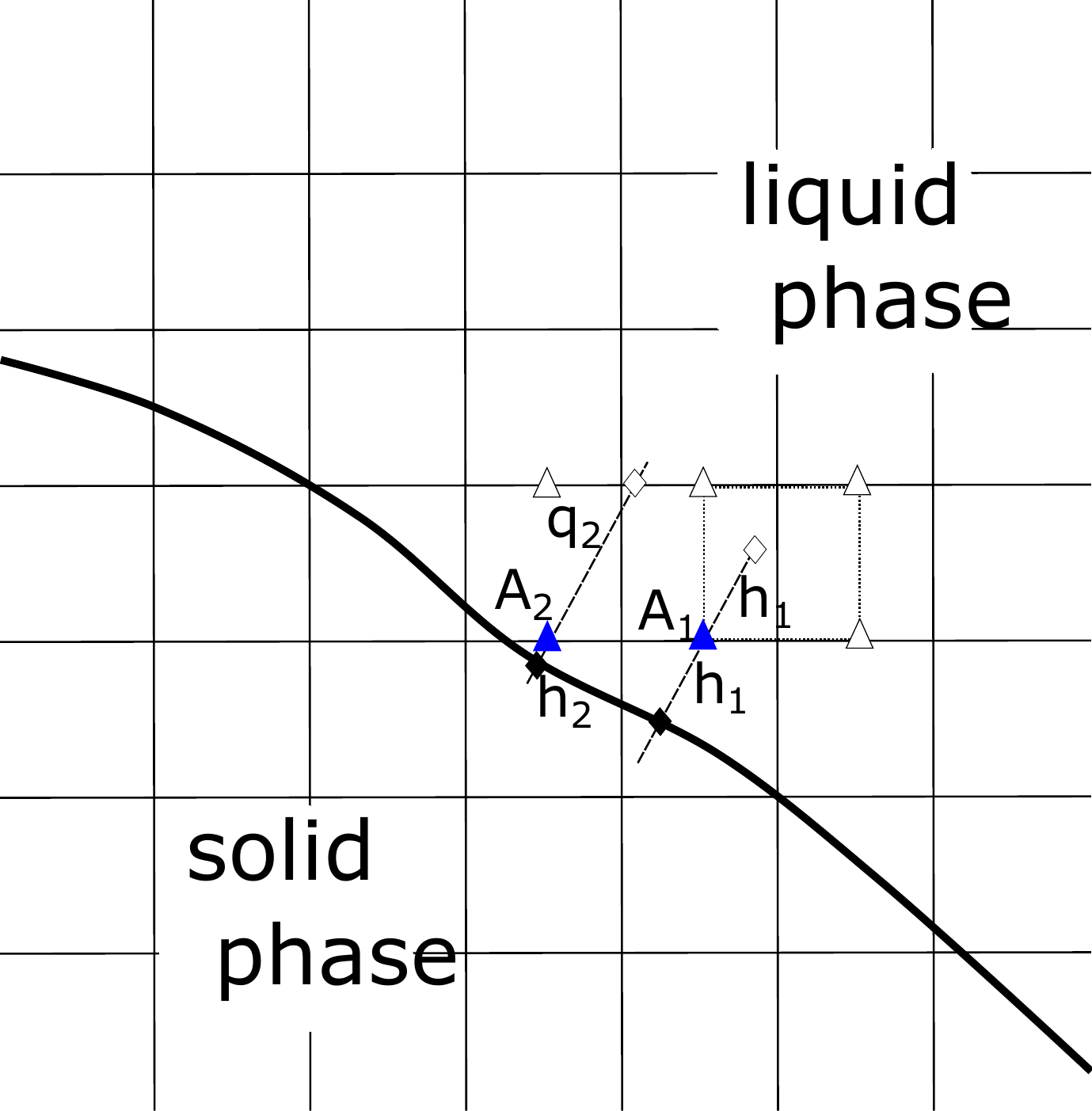}
    \caption{\label{fig:interpolation}}
  \end{subfigure}
  \caption{a) Definition of point tags: solid triangles are velocity solid points, empty triangles are fluid velocity points, red and blue triangles are forcing points for the horizontal and vertical velocity, respectively; solid circles are solid pressure points and empty circles are fluid pressure points. b) Definition of the interpolation stencil: forcing point (blue triangles), auxiliary point (empty diamond), solid boundary point (black diamond).\label{fig:punti}}
\end{figure}

The interaction between the fluid and the solid phases is dealt by the 
Immersed Boundary method \cite{peskin1972}. This technique avoids the use of 
body conforming meshes by introducing the term $\vec{f}$ in the momentum equation 
\eqref{eqn:momentum} which enforces the fluid velocity to be equal to that of 
the solid (no--slip boundary condition). Essentially, the IBM consists of two 
steps: \emph{i)} evaluation of the forcing term $\vec{f}$; \emph{ii)} evaluation 
of the hydrodynamic loads \eqref{eqn:loads}. Several methods have been proposed 
to implement these two steps, see for example 
\cite{Fadlun2000,Balaras2004,uhlmann2005,Vanella2009,Breugem2012,DeTullio2016}. 
In the direct forcing approach \cite{Fadlun2000}, the term $\vec{f}$ is computed 
by imposing that the velocity $\vec{u}^*$ matches the solid velocity $\vec{V}$ 
at the solid nodes
\begin{equation}
  \frac{\vec{V}^{l+1}-\vec{u}^l}{\Delta t} = -\frac{1}{\rho^{l+1}}\nabla p^l - \left(\frac{1}{\rho^{l+1}}-\frac{1}{\rho_0}\right)\nabla \hat{p} + \frac{3}{2}\vec{H}^l - \frac{1}{2}\vec{H}^{l-1} + \vec{g}^l + \vec{f}^l
\end{equation}
which yields for the force
\begin{equation}
 \vec{f}^l = \frac{\vec{V}^{l+1}-\vec{u}^l}{\Delta t} + \frac{1}{\rho^{l+1}}\nabla p^l + \left(\frac{1}{\rho^{l+1}}-\frac{1}{\rho_0}\right)\nabla \hat{p} - \frac{3}{2}\vec{H}^l + \frac{1}{2}\vec{H}^{l-1} - \vec{g}^l
\end{equation}
with the velocity $\vec{V}^{l+1}$ computed by \eqref{eqn:veltra} and \eqref{eqn:velrot}.
In practice this corresponds to locally reconstructing the velocity field at the
nodes next to the solid boundary (forcing nodes in figure \ref{fig:forcing}). 
It is worth noticing that, after the correction step \eqref{eqn:newcorrection}, 
which enforces the free divergence of the velocity field, the no--slip boundary 
condition might be violated. However, it has been shown that, provided the first
external node is within the linear part of the boundary layer profile, the error 
is small and it does not affect significantly the solution \cite{Fadlun2000}. 
It could be anyway further reduced by iterating between equations \eqref{eqn:veltra}, 
\eqref{eqn:velrot} and \eqref{eqn:projection}-\eqref{eqn:newcorrection} until 
the desired convergence is achieved.

Since the Cartesian grid does not conform to the solid body, the forcing points 
do not lie exactly at the boundary, as shown in figure \ref{fig:forcing}. 
For this reason a geometrical description of the solid is required to identify 
the forcing points and to interpolate the velocity in these points. 
The proposed scheme has been applied to bodies whose surface could be described 
by an analytical function since this allows a significant reduction of the 
computational cost. It is worth mentioning, however, that the method could be 
easily extended to general complex geometries by using a ray tracing procedure as
described in \cite{Iacca}. 

If, as in the present case, the solid body is a cylinder or a sphere, the geometry 
is simply defined by the position of the center and a distance from it. 
We introduce a signed distance function $d$ (positive in the fluid region and 
negative in the solid) from the surface of the body for every computational node.
In the same way we can define the local outward normal, by differentiation of 
the distance function $\vec{n} = \nabla d$. Starting from the distance field we can 
tag all the computational velocity points as follows: if $d < 0$ the point is solid 
and tagged with a $0$ flag (black triangles in figure \ref{fig:forcing}). Points 
where $d \ge 0$ and at least one neighbor with a negative distance are tagged as 
interface points with a flag $1$ (red and blue triangles in figure \ref{fig:forcing}); 
all other points with $d > 0$ are tagged as fluid with a flag $2$ (empty triangles 
in figure \ref{fig:forcing}). Pressure points, at the cell center, are tagged only 
as solid or fluid without the interface flag. For all the interface points 
the velocity needs to be locally reconstructed in order to apply the no--slip 
boundary condition. This is done, similarly to \cite{Balaras2004}, by interpolation 
between the velocity value on the solid body and a virtual point in the fluid domain, 
as shown in figure \ref{fig:interpolation}. Starting from the forcing node, 
following the local normal, a support of $4$ total adjacent points ($8$ in 3D) 
is constructed. If all the points, except the forcing one, are fluid (as for $A_1$ 
in figure \ref{fig:forcing}) then the auxiliary point is located at a distance $h$ 
equal to the distance between the forcing node and the solid boundary. The velocity 
is computed by bi--linear interpolation at the auxiliary point and the desired 
velocity at the forcing node by linear interpolation between this auxiliary and 
the boundary nodes (black diamond in figure \ref{fig:forcing}). In case one of 
the adjacent points is another forcing node (for example $A_2$ has as neighbor $A_1$), 
then the auxiliary point is found by intersection of the local normal and the 
Cartesian grid. The velocity here is computed using the adjacent points and then 
the velocity on the forcing point is evaluated as done for the previous case. 
This method provides the value of the velocity $\vec{V}^{l+1}$ on the forcing node 
to impose the desired boundary condition, and it avoids the inversion of a matrix, 
as proposed in \cite{Yang2006}. Note that the interpolation is performed only for 
the interface points, while for solid ones the velocity is imposed directly
as the solid body velocity $\vec{V}^{l+1}$.

\subsection{Hydrodynamic forces and solid body dynamics}

The dynamics of the solid phase is governed by the hydrodynamic forces 
\eqref{eqn:loads} which are computed by integrating over the body surface the 
integral of viscous stress and pressure. Given that velocity gradients and pressure 
are not defined exactly at the immersed surface also in this case interpolation
is necessary.
This is accomplished by considering virtual probes in the fluid domain where $\tau$ 
and $p$ are computed and then projected along the normal direction onto the surface. 
Provided the first external point is within the linear part of the boundary layer
velocity profile, the viscous stress on the surface is then equal to the value at 
the probe. Instead, the pressure on the surface is evaluated imposing the boundary 
condition
\begin{equation}
  \frac{\partial p}{\partial n} = -\rho \frac{D\vec{u}}{Dt}\cdot\vec{n} + \rho \vec{g}\cdot\vec{n},
\end{equation}
which is obtained by projecting the momentum equation in the wall normal direction.
The surface pressure is then
\begin{equation}
  p = p_s + h_s\left(\rho \frac{D\vec{u}}{Dt}\cdot\vec{n} - \rho \vec{g}\cdot\vec{n}\right),
\end{equation}
being $p_s$ the pressure at the probe and $h_s$ the distance between the probe 
and the immersed surface. The probes are initially located at a distance $\Delta$ 
from the surface and a support of four nodes (eight in 3D) , 
as in figure \ref{fig:interpolation}, 
is built in the wall normal direction. If these points are all fluid nodes then 
$\tau$ and $p$ are computed in the probe by bilinear interpolation, otherwise the 
probe distance is progressively increased by steps of $0.2\Delta$ until the condition 
on the support is verified. The number of probes is computed by considering a spacing 
equal to $\Delta$. Once the forces are known, equations 
\eqref{eqn:veltra}-\eqref{eqn:posrot} are advanced in time using a Hamming 
$4^{th}$ order modified predictor--corrector method \cite{Hamming1959}, described in 
appendix \ref{sec:A}.

\subsection{Summary of the algorithm}
To summarize the algorithm let's define
$\vec{\mathcal{X}} = \left[\vec{X}, \vec{\Theta}\right]^T$ and $\vec{\mathcal{V}} = \left[\vec{V}, \vec{\Omega}\right]^T$ the generalized position and velocity
for each solid body; for every timestep (after the third), starting 
from $\vec{u}^l$, $p^l$, $\mathcal{F}^l$, $\vec{\mathcal{X}}^l$ and $\vec{\mathcal{V}}^l$, the procedure is the following:
\begin{enumerate}
\item compute the predicted imporved solution $\left(\vec{\mathcal{X}_m}^{l+1},
\vec{\mathcal{V}_m}^{l+1}\right)$;
\item solve equation \eqref{eqn:vof} to compute $\mathcal{F}^{l+1}$;
\item update fluid properties with \eqref{eqn:properties};
\item compute the provisional velocity field \eqref{eqn:projection};
\item reconstruct the velocity field to impose the no--slip boundary condition 
given by $\left(\vec{\mathcal{X}_m}^{l+1},\vec{\mathcal{V}_m}^{l+1}\right)$ on the solid boundary;
\item solve Poisson equation \eqref{eqn:poisson} to compute $\phi^l$;
\item compute $p^{l+1}$ by \eqref{eqn:newp} and $\vec{u}^{l+1}$ by \eqref{eqn:newcorrection};
\item compute the hydrodynamic loads $\vec{F}^l_1$ and $\vec{M}^l_1$ by \eqref{eqn:loads};
\item perform the correction step of the Hamming solver and compute ;
$\left(\vec{\mathcal{X}_1}^{l+1},\vec{\mathcal{V}_1}^{l+1}\right)$;
\item repeat steps from 2. to 9. with boundary condition $\left(\vec{\mathcal{X}_1}^{l+1},\vec{\mathcal{V}_1}^{l+1}\right)$ for step 5. and check for convergence on $\left(\vec{\mathcal{X}_k}^{l+1},\vec{\mathcal{V}_k}^{l+1}\right)$;
\item if not converged continue to iterate over steps 2. to 9.;
\item if converged, compute the final solution $\left(\vec{\mathcal{X}}^{l+1},\vec{\mathcal{V}}^{l+1}\right)$ and repeat steps 2. to 8. with this boundary condition for the velocity field.
\end{enumerate}
Note that the strong coupling (\emph{i.e.} the iterative solver) is necessary only for large density ratio between fluid and solid; in other cases it is possible to use a direct solver performing steps 1-8. When using the strong coupling, the first three timesteps are performed with a forward Euler, Adam-Bashfort 2nd order and Adam-Bashfort 3rd order method, respectively; a tolerance of $10^{-6}$ for convergence is usually small enough for an accurate solution. In case of large acceleration
under-relaxation can be used to stabilize the simulation by replacing the force in the corrector step with $F = (1-\chi)F_k + \chi F_{k-1}$, with $\chi = [0.1, 0.3]$ \cite{deTullio2009}. For the descirption of the Hamming's method see appendix A.

\section{Results}\label{sec:results}

\subsection{Validations}\label{sec:validations}

We present here some classical tests either for the immersed boundary method and 
for its interaction with an interface separating two fluids. We also verify that 
the method is able to compute the proper energy decay of surface gravity waves
and the oscillation period of a submerged pendulum. Note that in all simulations 
we have set surface tension coefficient to zero to avoid further restrictions on the 
timestep.

\subsubsection{Particle migration}

\begin{figure}
  \centering
  \begin{subfigure}{0.45\textwidth}
    \includegraphics[width=\textwidth]{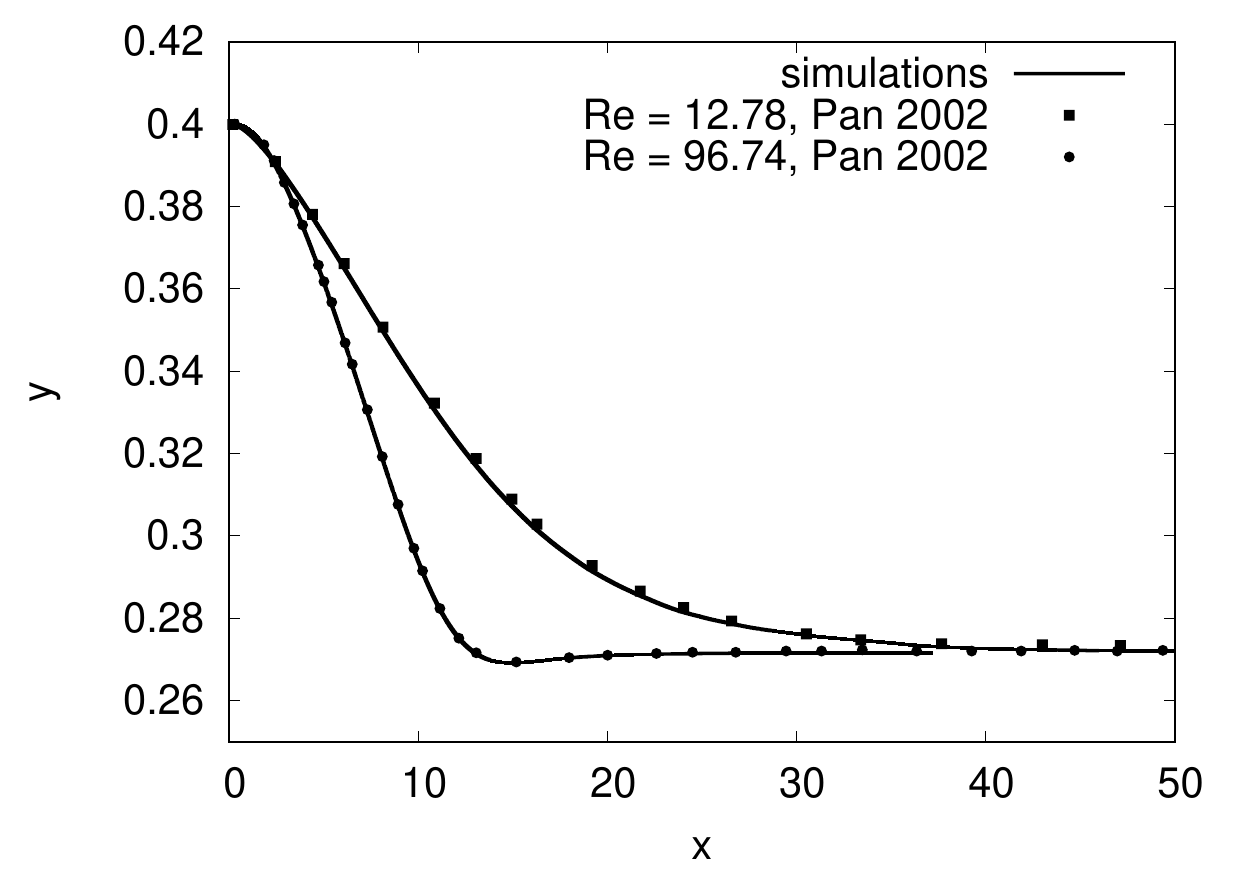}
    \caption{\label{fig:Pancomp}}
  \end{subfigure}
  \begin{subfigure}{0.45\textwidth}
    \includegraphics[width=\textwidth]{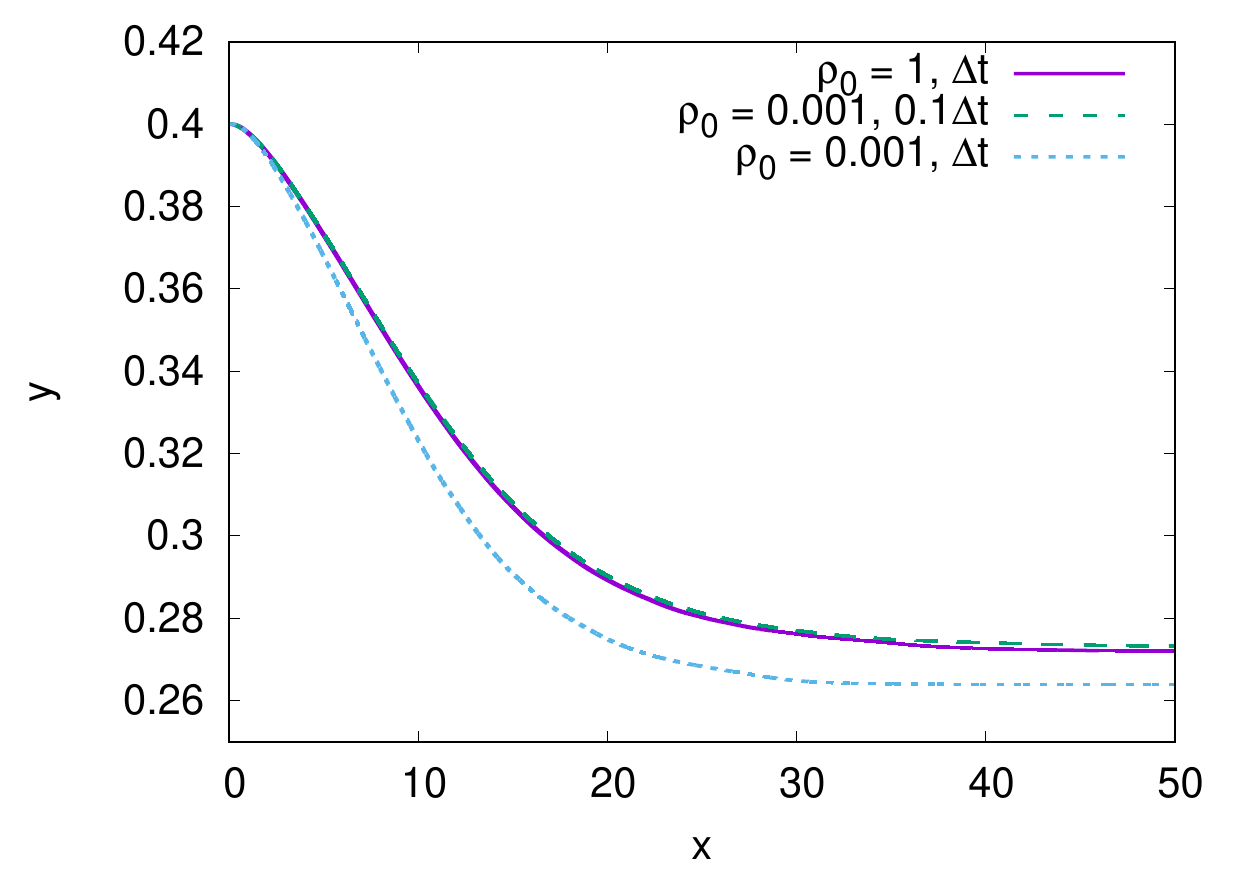}
    \caption{\label{fig:Pandt}}
  \end{subfigure}
  \caption{Lateral migration of a circular particle in a pressure--driven channel
 at $Re = 12.78$ and $96.74$: (a) vertical $vs$ horizontal position; solid line 
for the present results and symbols for \cite{Pan2002}. (b) Effect of the splitting 
\eqref{eqn:splitting}: case without splitting (solid purple line), case with splitting 
and $\Delta t$ ten times smaller (dashed green line), case with splitting and 
the same $\Delta t$ (dotted blue line).\label{fig:Pan}}
\end{figure}

In the first test case we consider the migration of a neutrally buoyant particle 
in a two--dimensional pressure--driven Poiseuille flow \cite{Pan2002}. The motion 
of the particle is a consequence of pressure distribution, inertia and rotational lift;
an accurate evaluation of the forces is necessary to properly simulate the particle 
dynamics. We consider a square domain of size $L = 1$, periodic in the horizontal 
direction and vertically bounded by two solid walls. Initially the flow is at rest 
and a pressure gradient is applied in the horizontal direction. The circular
particle has a radius of $0.125L$ and it is initially located at a vertical distance
of $0.4L$ from the lower plate. Two different cases are considered by varying the 
viscosity of the fluid $\mu$, which correspond to two different Reynolds numbers, 
namely $Re = 12.78$ and $96.74$. The Reynolds number is based on the space--averaged 
inlet velocity $\tilde{u}$ and the size of the domain $L = 1$, \emph{i.e.} 
$Re = \rho L\tilde{u}/\mu$, the density of the particle $\rho_p$ is the same as
that of the fluid $\rho$. For the lower value of $Re$ we use a grid of $96\times96$ 
points while for the higher $Re$ the grid is $192\times192$. The particle migrates 
from the initial to an equilibrium position close to the lower wall, as shown in 
figure \ref{fig:Pancomp}. A good agreement for both Reynolds numbers, with the 
results of \cite{Pan2002}, is found. 

To show the effect of the splitting procedure \eqref{eqn:splitting} we simulate 
the case at $Re = 12.78$ by introducing a minimum density equal to $\rho_0 = 0.001$, 
which corresponds to a case with two fluids with density ratio 0.001, so that the 
term involving $\hat{p}$ is non-zero. Due to the small value of $\rho_0$, it is 
necessary to decrease the timestep in order to recover the approximation between 
$\hat{p}$ and $p^{l+1}$, as clearly shown in figure \ref{fig:Pandt}. The timestep 
restriction depends on the density ratio and we have found that a reduction of $10$
times for a density ratio of $0.001$ is enough while for bigger density ratios, 
for example $0.01$, already halving the timestep yields the correct time evolution 
of the particle. This is consistent with the results of \cite{Frantzis2019}. For this test
the direct solver is stable and it is not necessary to use the iterative solver.

\subsubsection{Water exit}

\begin{figure}
  \centering
  \begin{subfigure}{0.3\textwidth}
    \includegraphics[width=\textwidth]{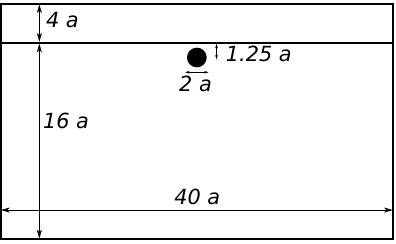}
    \caption{\label{fig:WaterExitSetch}}
  \end{subfigure}
  \begin{subfigure}{0.3\textwidth}
    \includegraphics[width=\textwidth]{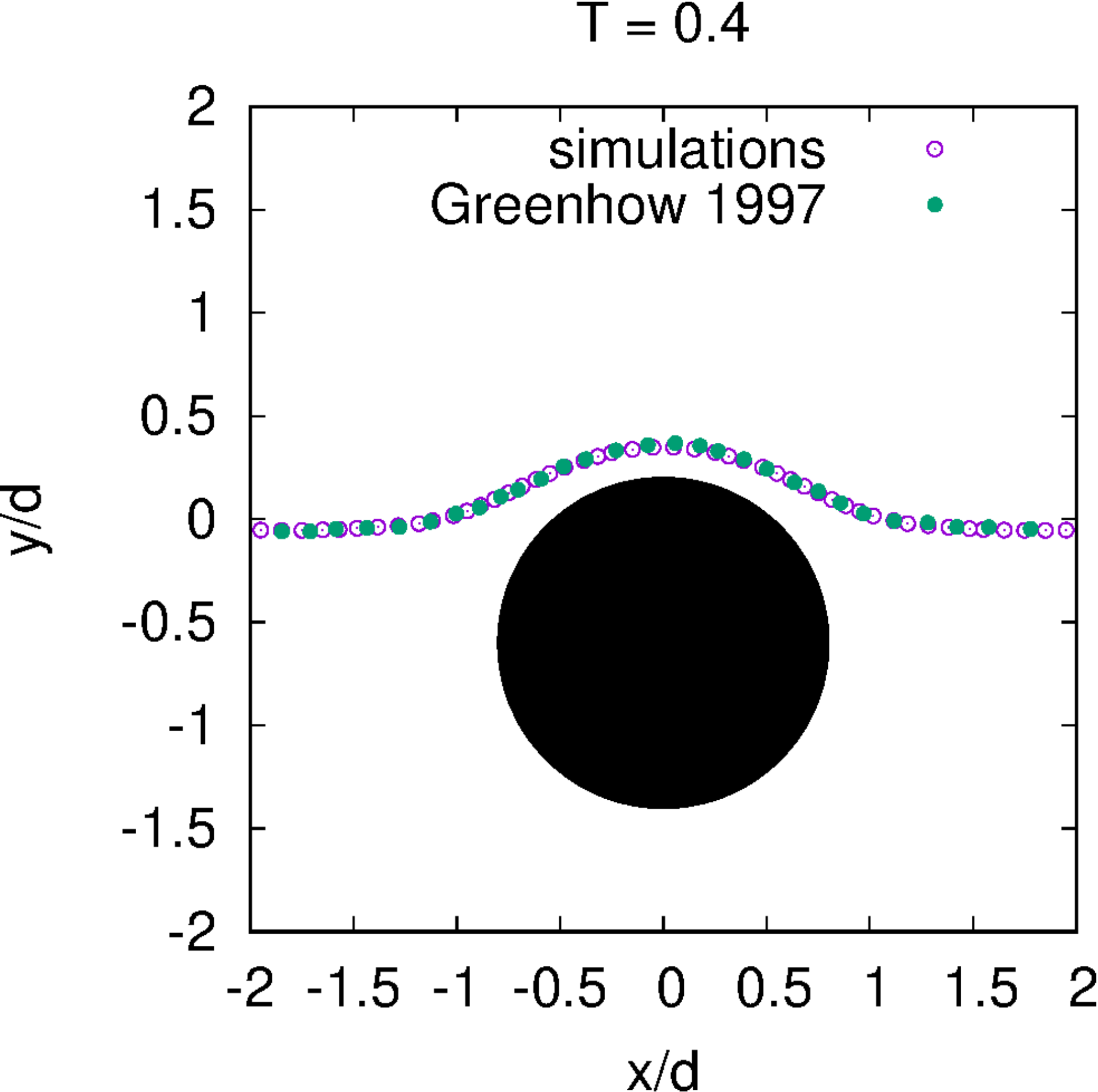}
    \caption{\label{fig:Lin04}}
  \end{subfigure}
  \begin{subfigure}{0.3\textwidth}
    \includegraphics[width=\textwidth]{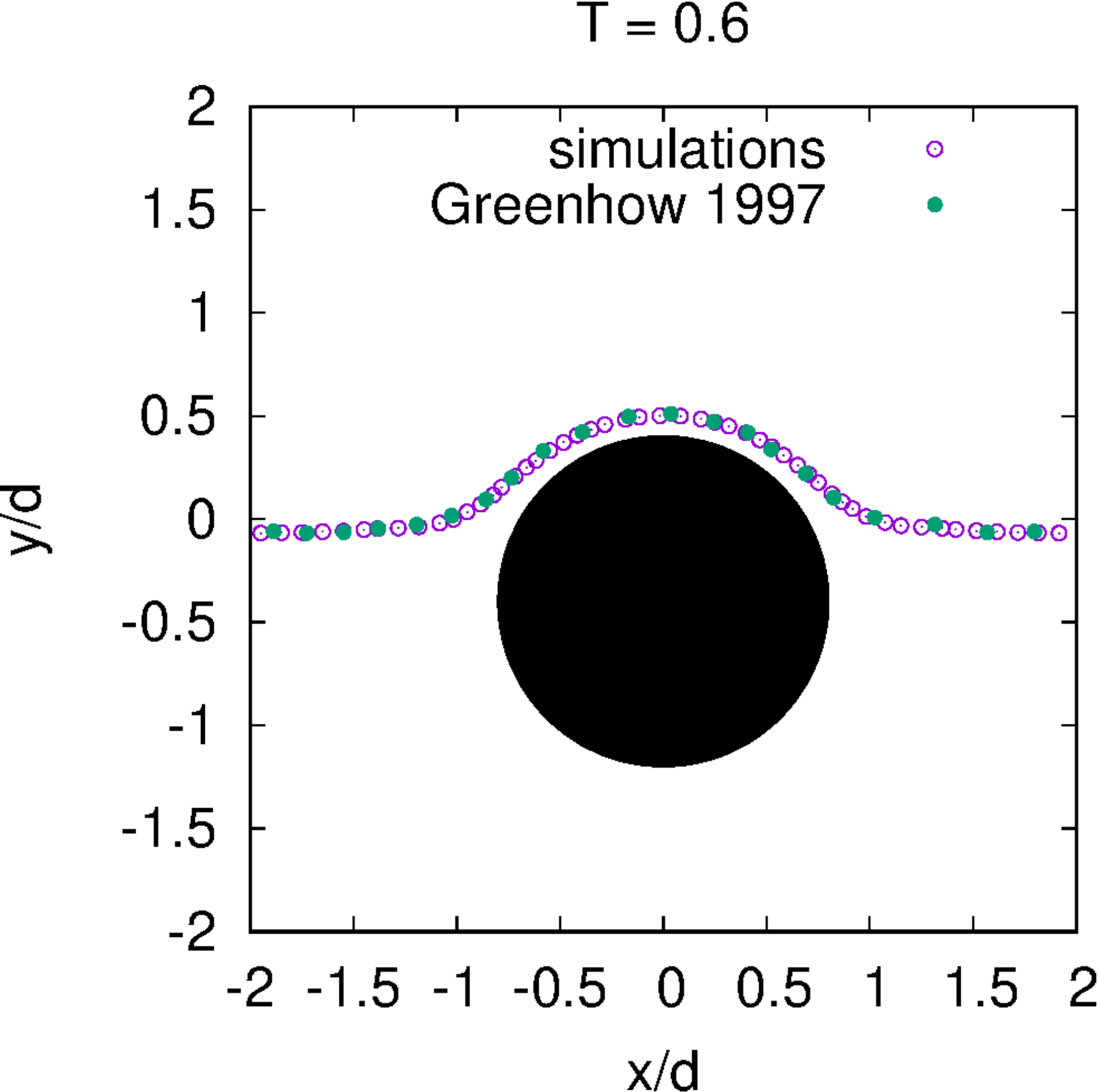}
    \caption{\label{fig:Lin06}}
  \end{subfigure}
  \caption{Water exit test case: a) sketch of the computational domain; b)-c) interface deformation with $\epsilon = 0.8$ and $Fr = 0.39$: present results (open circles), data from \cite{Greenhow1997} (solid circles); lengths are made non--dimensional by 
the initial distance $d$.\label{fig:Greenhow}}
\end{figure}

The next test case consists of a cylinder rising close to the interface separating 
water and air, as in \cite{Greenhow1997}. The cylinder has radius $a$, it is 
at a distance $d$ below the still water level, and rises with a constant velocity $V$. 
The inviscid problem is governed by two nondimensional parameters, 
$\epsilon = a/d = 0.8$ and 
the Froude number $Fr = V/\sqrt{gd} = 0.39$, with $g$ the acceleration of gravity. 
The material properties ratio is that of water and air, 
\emph{i.e.} $\rho_w/\rho_a = 850$ and $\mu_w / \mu_a = 1.92\text{x}10^{-2}$. 
Note that the reference solution of 
\cite{Greenhow1997} has been derived from the potential theory for inviscid fluids; 
the present solver, instead, is designed for viscous flows therefore we have selected 
a Reynolds number high enough to minimize viscous effects but, at the same time,
small enough to limit the computational effort. In the real air/water case it results
$Re \approx {\cal O}(10^5)$; here we have set the water viscosity so to have
a Reynolds number $Re = \rho_w V a / \mu_w = 1000$ which proved to be enough for our
purposes. 
The domain is $40 a$ wide and $20 a$ high, with a water depth of $16 a$ and it is 
discretized by a grid of $4096\times2048$ points. We compare our results with the 
numerical ones proposed by \cite{Greenhow1997}, reported in figure \ref{fig:Greenhow} 
at fixed non--dimensional times of $T = tV/d = 0.4$ and $0.6$. The results, 
in terms of interface deformation, are in very good agreement with those of
\cite{Greenhow1997}. Also for this test the direct solver is enough stable and it is not necessary 
to use the iterative solver.

\subsubsection{Viscous damping of surface gravity waves}

\begin{figure}
  \centering
  \begin{subfigure}{0.49\textwidth}
    \includegraphics[width=0.8\textwidth]{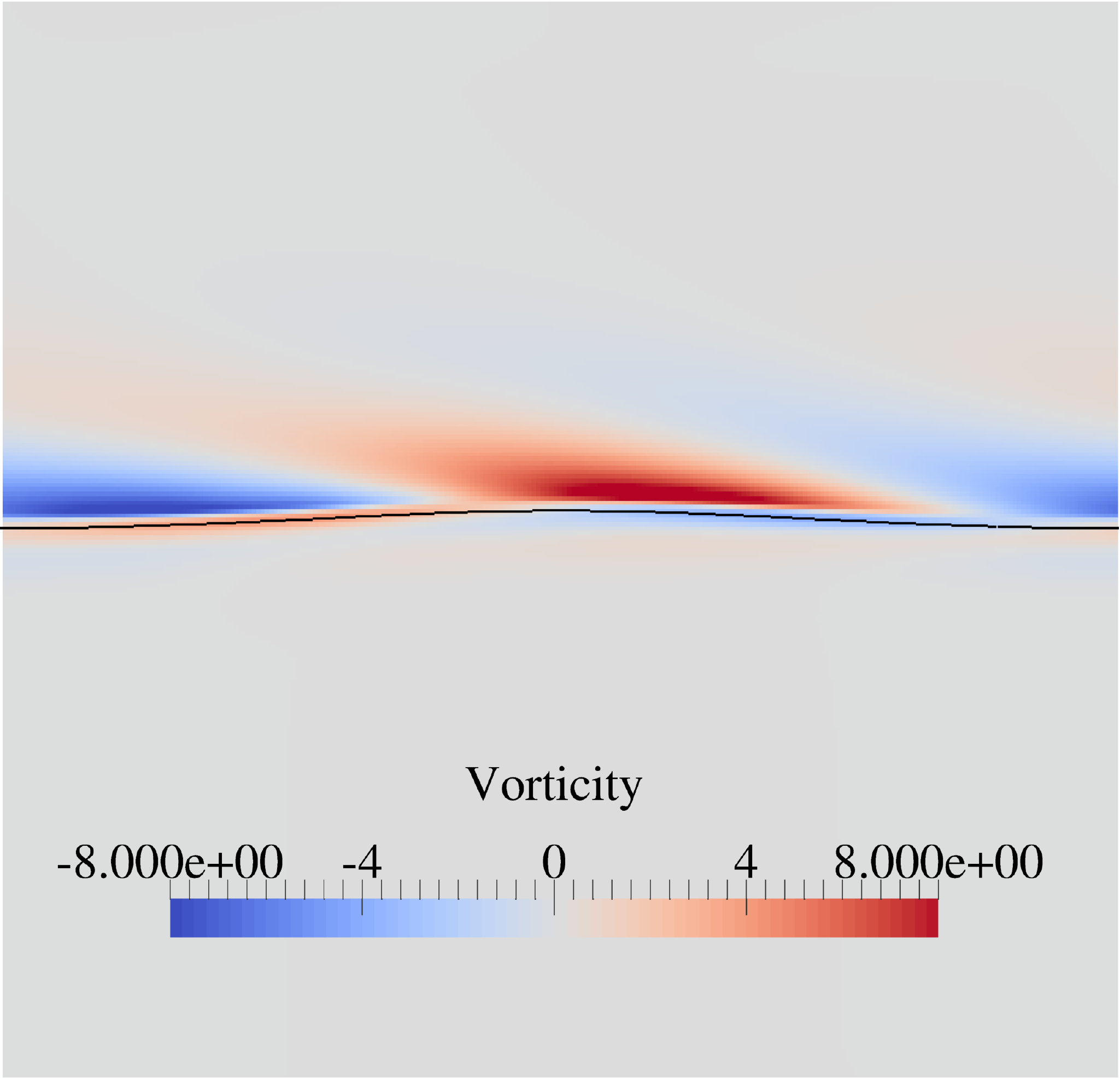}
  \end{subfigure}
  \begin{subfigure}{0.49\textwidth}
    \includegraphics[width=\textwidth]{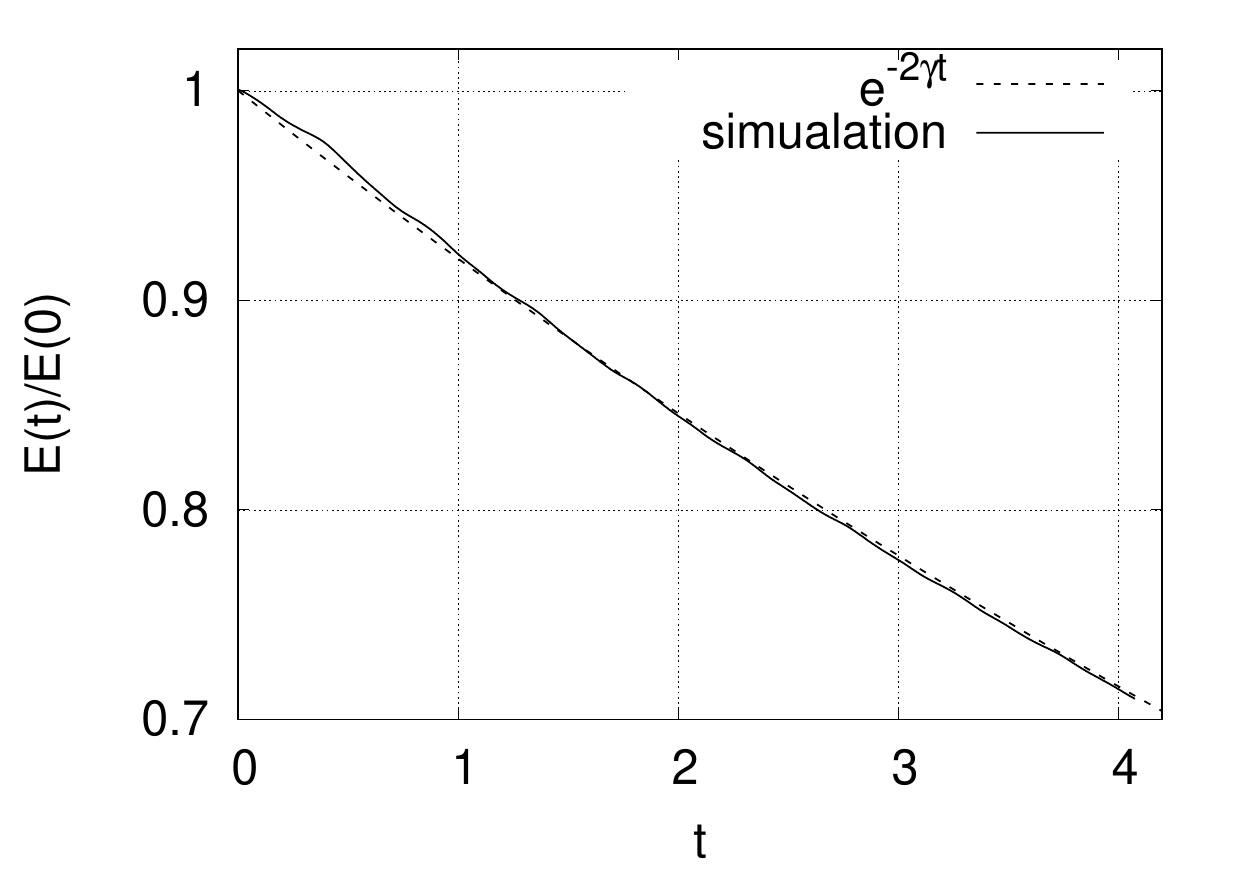}
  \end{subfigure}
  \caption{(Left panel) Vorticity contours after one wave period; (Right panel) 
Comparison of the energy decay between the computed values from
 equations \eqref{eqn:kin}-\eqref{eqn:pot} and the theoretical 
exponential decay.\label{fig:decay}}
\end{figure}

Surface gravity waves propagating in inviscid fluids can be described using the 
potential theory which yields an irrotational velocity field. For viscous fluids, 
instead, there is an additional rotational flow given by the presence of the 
boundary layer at the free--surface. For waves of small amplitude, \emph{i.e.} 
waves for which $a << \lambda$, with $a$ the wave amplitude and $\lambda$ its
wavelength, the rotational flow is confined in a small layer across the surface 
while the rest of the flow can be still described by the potential theory. Under 
this assumption the total wave energy decay in time, given by the sum of the 
kinetic and potential contributions, can be computed from the potential theory 
(see \citeauthor{Landau} \cite{Landau} chapter II section 25) and it results in 
an exponential decay of the form $E(t) = E(t=0)e^{-2\gamma t}$ with the 
coefficient $\gamma = 2 \nu k^2$, being $\nu$ the kinematic viscosity of the 
fluid and $k = 2 \pi /\lambda$ the wavenumber. The kinetic energy contribution 
is given by
\begin{equation}
  \label{eqn:kin}
    E_k(t) = \int_0^{\lambda} \int_{-h}^{\eta} \frac{1}{2}\rho \vec{u}^2\,dxdy
\end{equation}
with $h$ the fluid depth at rest and $\eta$ the free--surface level
with respect to the equilibrium position ($z=0$). The potential energy is 
\begin{equation}
  \label{eqn:pot}
    E_p(t) = \int_0^{\lambda} \int_{-h}^{\eta} \rho gz\,dxdy - E_{p0}
\end{equation}
with $E_{p0} = \int_0^{\lambda} \int_{-h}^{0} \rho gz\,dxdy = -\rho g\lambda h^2/2$ 
the potential energy of the still water level. We simulate the propagation of a 
surface gravity wave of wavelength $\lambda$ in a periodic square box of size 
$\lambda$ with a fluid depth of $h = \lambda / 2$ and a steepness 
$\varepsilon = a k = 0.05$. The density and viscosity ratios are those of water 
and air, $1/850$ and $1.96 \text{x} 10^2$, respectively, while the Reynolds number 
based on the phase velocity $c_f = \sqrt{g \lambda}$ and on the wavelength 
$\lambda$ is $Re = g^{0.5} \lambda^{3/2} / \nu = 1.0 \text{x} 10^4 $. The initial 
conditions for the surface elevation $\eta$ and the velocity field $\vec{u} = (u,w)$ 
are given by the linear theory:
\begin{subequations}
  \begin{align}
    \eta(x,0) &= a \cos\left(kx\right), \\
    u(x,y,0) &= \left(1-\mathcal{H}\right)a\omega e^{ky} \cos\left(kx\right) - \mathcal{H}a\omega e^{-ky} \cos\left(kx\right), \\
    v(x,y,0) &= \left(1-\mathcal{H}\right)a\omega e^{ky} \sin\left(kx\right) + \mathcal{H}a\omega e^{-ky} \sin\left(kx\right),
  \end{align}
\end{subequations}
with $\mathcal{H} = 0$ for the water and $\mathcal{H} = 1$ in the air. For stability 
reason the initial VoF function $\mathcal{H}$ is filtered one time using bilinear 
interpolation. In figure \ref{fig:decay} (left) we report the vorticity contours 
after one period, which clearly shows its concentration in a small layer below the 
water surface; in the right panel we show the energy decay in time computed by
 equations \eqref{eqn:kin}-\eqref{eqn:pot} alongside the theoretical exponential 
decay with excellent agreement. The simulation has been performed on a grid of 
$256\times256$ points.

\subsubsection{Frequency oscillations of a submerged reversed pendulum}

\begin{figure}
  \centering
  \begin{subfigure}{0.49\textwidth}
    \includegraphics[width=\textwidth]{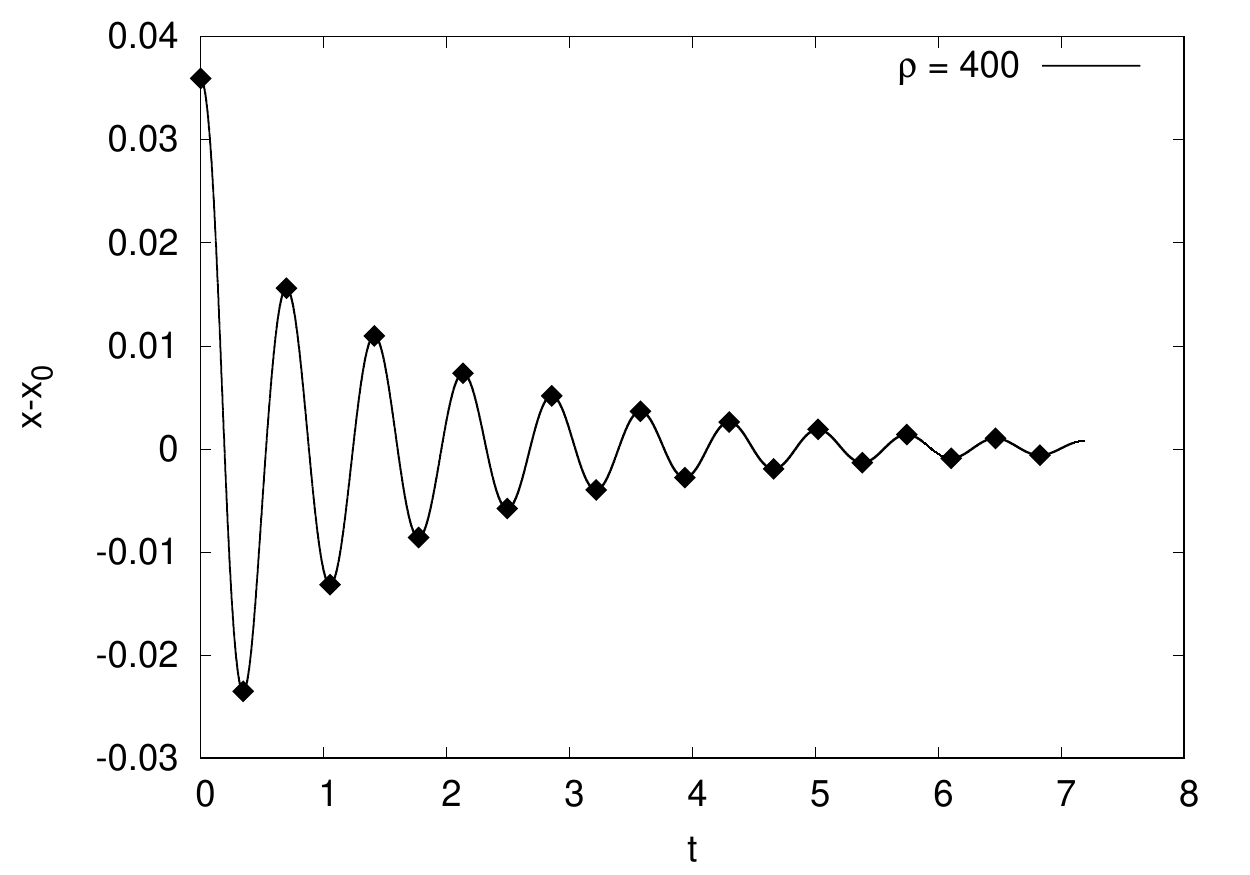}
  \end{subfigure}
  \begin{subfigure}{0.49\textwidth}
    \includegraphics[width=\textwidth]{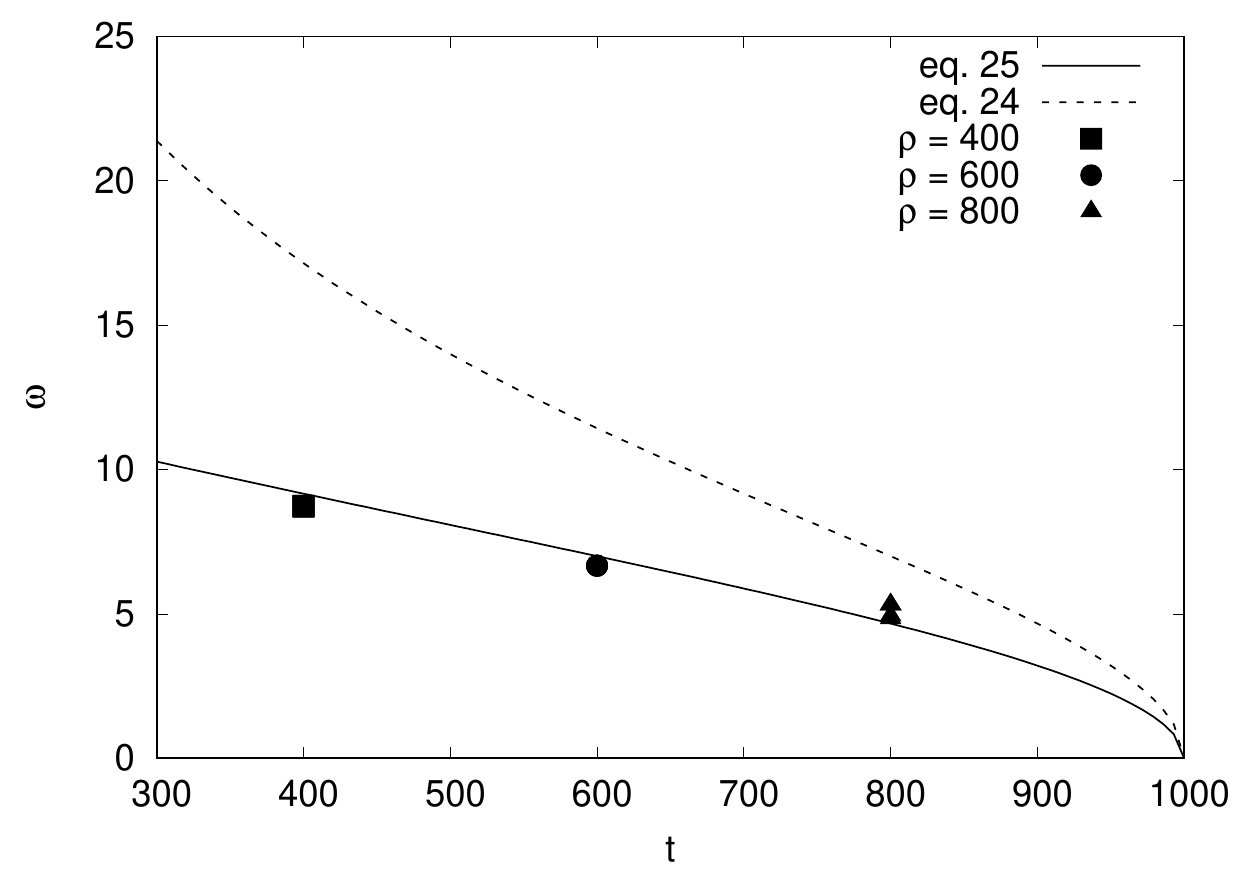}
  \end{subfigure}
  \caption{Oscillation frequency of a submerged reversed pendulum: (left) time 
history of the horizontal position of the center of mass for one case with 
$\rho = 400$, symbols highlight the maxima and minima used for the computation 
of the period; (right) comparison of the computed frequency with equations 
\eqref{eqn:buoy} and \eqref{eqn:added}. \label{fig:frequency}}
\end{figure}

We considere here a 2D reversed pendulum, \emph{i.e.} a cylinder lighter than ther surrounding fluid and anchored at the bottom. The natural frequency of a simple pendulum, for small oscillation amplitude, is 
given by $\omega = \sqrt{g/\ell}$, with $g$ the gravity and $\ell$ the distance of the 
center of rotation from the center of mass, and is obtained by a balance between 
the tension of the contraint and the weight. On account also of the buoyancy force 
the frequency becomes
\begin{equation}
  \label{eqn:buoy}
  \omega = \sqrt{\frac{g(\rho-\rho_p)}{\ell \rho_p}},
\end{equation}
where $\rho$ is the density of the surrounding fluid and $\rho_p$ that of the pendulum.
This relation is appropriate when the pendulum is much denser than the fluid whereas 
in the general case it is necessary to take into account also the added mass:
\begin{equation}
  \label{eqn:added}
  \omega = \sqrt{\frac{g}{\ell}\frac{\rho-\rho_p}{\rho_p + c\rho}}
\end{equation}
with $c$ the added mass coefficient depending on the shape of the swinging mass 
(for example $c=0.5$ for spheres and $c=1$ for cylinders). Here we compute the 
oscillation frequency for a cylinder (2D case) for different values of the density 
$\rho$ and compare the results with equations \eqref{eqn:buoy} and \eqref{eqn:added}. 
To this end, we give an initial displacement to the center of mass of the cylinder 
from the equilibrium position and let the pendulum oscillate; after $10$ periods we 
compute the time difference between all maxima and minima of the horizontal position 
of the pendulum (as in figure \ref{fig:frequency}). In figure \ref{fig:frequency}b 
we compare the computed frequency of the pendulum with equations \eqref{eqn:buoy} 
and \eqref{eqn:added}: the results show that including the added mass term gives a 
good approximation of the evaluations of the pendulum frequency, with an error 
about 5\%. 

The simulations are performed with the following setup: radius of the pendulum 
$r = 1$, fluid density $\rho = 1000$, fluid viscosity $\mu = 2.6\text{x}10^{-3}$, 
pendulum length $\ell = 1.8r$, size of the domain $10r\text{x}10r$; the grid has 
$256\times256$ points. The iterative solver has been used with a tolerance of $10^{-6}$.

\subsection{Applications}

\subsubsection{Surface gravity wave propagating over a submerged reversed pendulum}

We study now the interaction between surface gravity waves propagating over a 
submerged reversed pendulum. The wave has wavelength $\lambda$ and propagates with 
a phase velocity $c_p$ from the left to the right. The pendulum of radius $r$, is 
located (at rest) at a distance $d$ from the still water level and it is anchored 
at a distance $\ell$ from its center of mass. The density of the pendulum $\rho_p$ 
is smaller than the water density $\rho_w$, the buoyancy pulls the cylinder upwards 
and the constraint is always under positive tension. From the linear theory, the 
frequency of the wave is
\begin{equation}
  \omega_w = \sqrt{gh\tanh{kh}},
\end{equation}
with $g$ the gravitational acceleration, $k = 2\pi/\lambda$ the wavenumber and $h$ 
the depth of the still water level. The frequency of the pendulum can be estimated 
using equation \eqref{eqn:added}.

The problem depends on several parameters, which make the analysis quite complicated; 
here we show one case with the following setup:

\begin{center}
  \begin{tabular}{ |c|c|c|c|c|c|c|c|c| }
    \hline
    $\lambda/r$ & $h/r$ & $\ell/r$ & $d/r$ & $\rho_p/\rho_w$ & $\rho_a/\rho_w$ & $\mu_a/\mu_w$ & $\varepsilon = ak$ & $Re$ \\
    \hline
    33.615 & 8.40375 & 2.5 & 1.65 & 0.8 & 1/850 & 1.96e-2 & 0.05 & $10^5$ \\ 
    \hline
  \end{tabular}
\end{center}
with the Reynolds number defined as $Re = \rho_w\lambda^{3/2}g^{1/2}/\mu_w$. 
Note that with this choice of the parameters the period of the pendulum is $1.6$
 times that of the wave. The domain is wide $\lambda$ and high $0.5\lambda$, the 
pendulum is located at the center of the domain in the horizontal direction; 
the initial wave profile is
\begin{equation}
  \eta(x,0) = a\cos(kx),
\end{equation}
while the initial velocity field in the water is 
\begin{subequations}
  \begin{align}
    &u(x,y,0) = a\omega\frac{\cosh\left(k\left(y+h\right)\right)}{\sinh\left(kh\right)}\cos\left(kx\right), \\
    &v(x,y,0) = a\omega\frac{\sinh\left(k\left(y+h\right)\right)}{\sinh\left(kh\right)}\sin\left(kx\right)
  \end{align}
\end{subequations}
and in the air the velocity has the same expression except for a change of sign in
 $y$. To ensure that the cord is always at the maximum extension, \emph{i.e.} $\ell$ 
is a rigid constraint; accordingly we compute the hydrodynamic forces and then solve 
for the rotation around the fulcrum of the pendulum; velocity and position of the center of mass are then computed from the angular quantities. Hence, there is no rotation of the pendulum around its center of mass.

\begin{figure}
  \centering
  \begin{subfigure}{0.49\textwidth}
    \includegraphics[width=\textwidth]{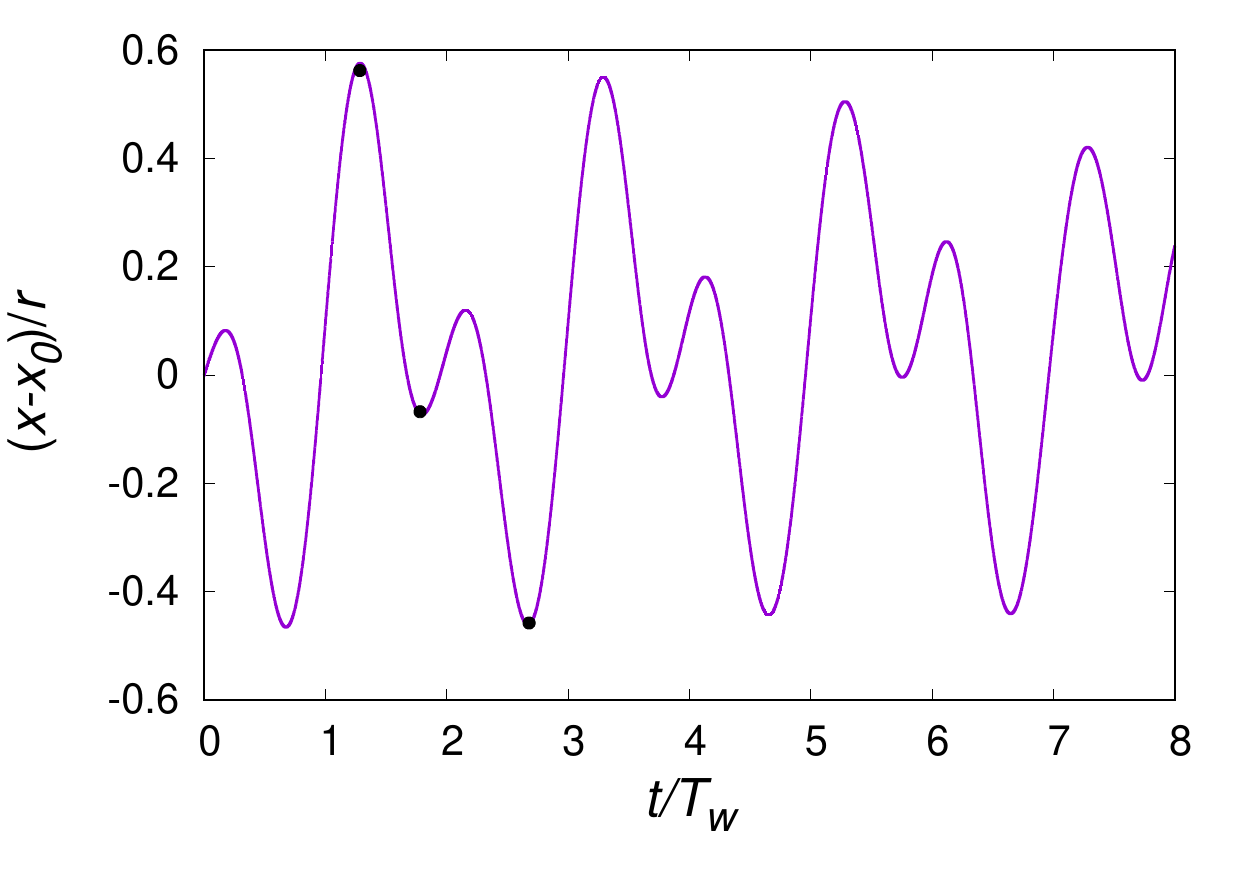}
    \caption{\label{fig:pendpos}}
  \end{subfigure}
  \begin{subfigure}{0.49\textwidth}
    \includegraphics[width=\textwidth]{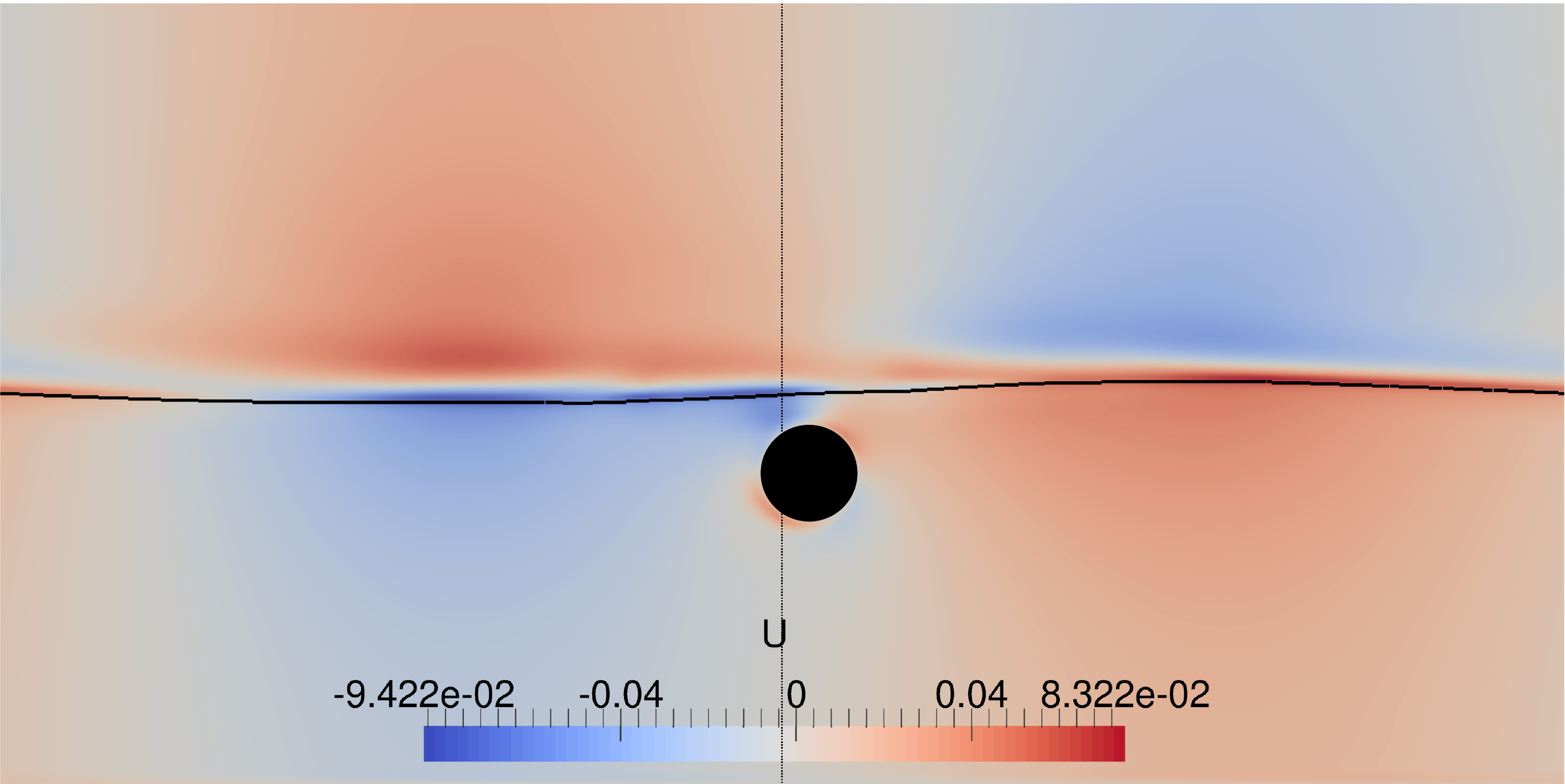}
    \caption{\label{fig:t1}}
  \end{subfigure}
  \begin{subfigure}{0.49\textwidth}
    \includegraphics[width=\textwidth]{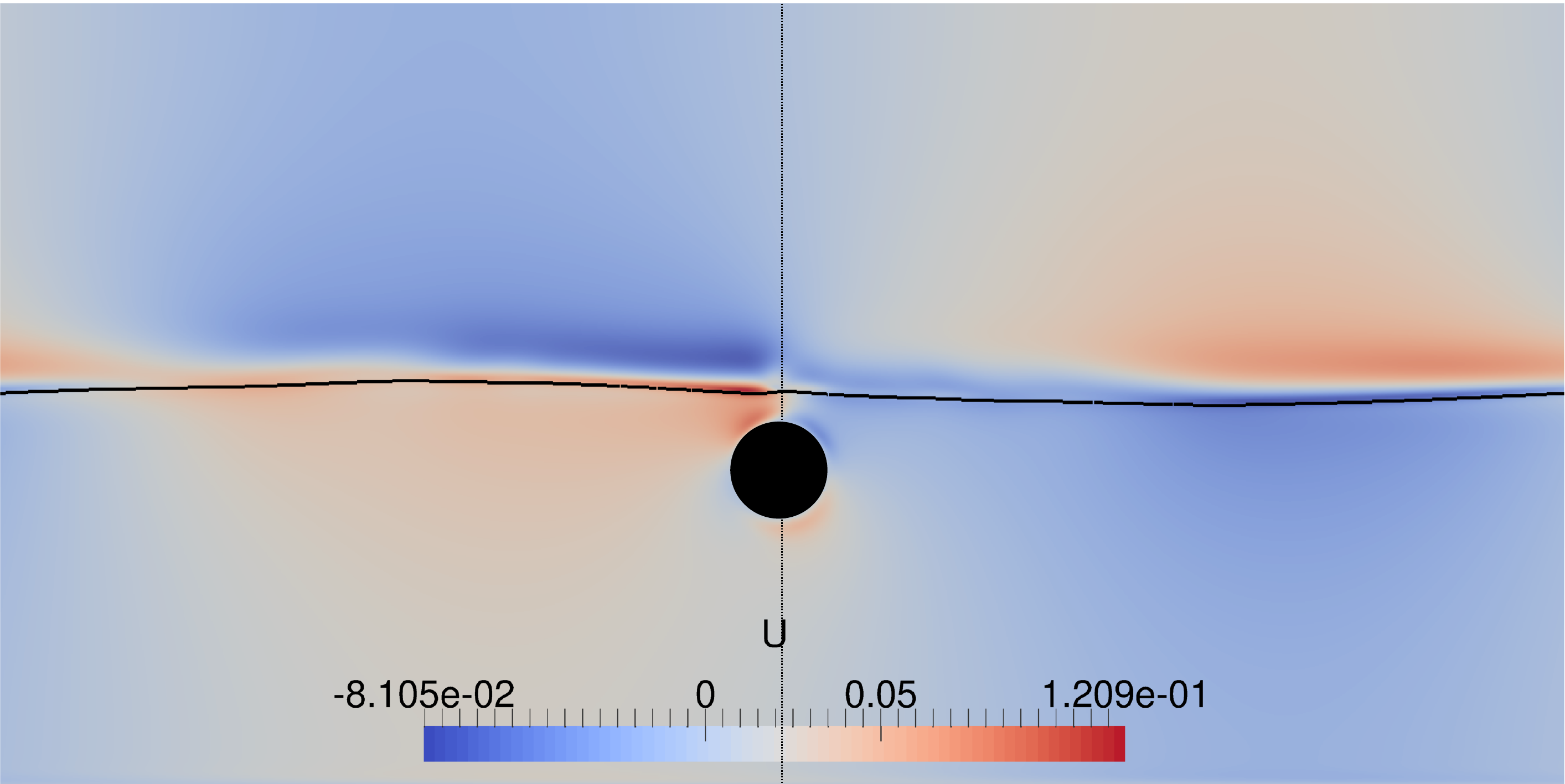}
    \caption{\label{fig:t2}}
  \end{subfigure}
  \begin{subfigure}{0.49\textwidth}
    \includegraphics[width=\textwidth]{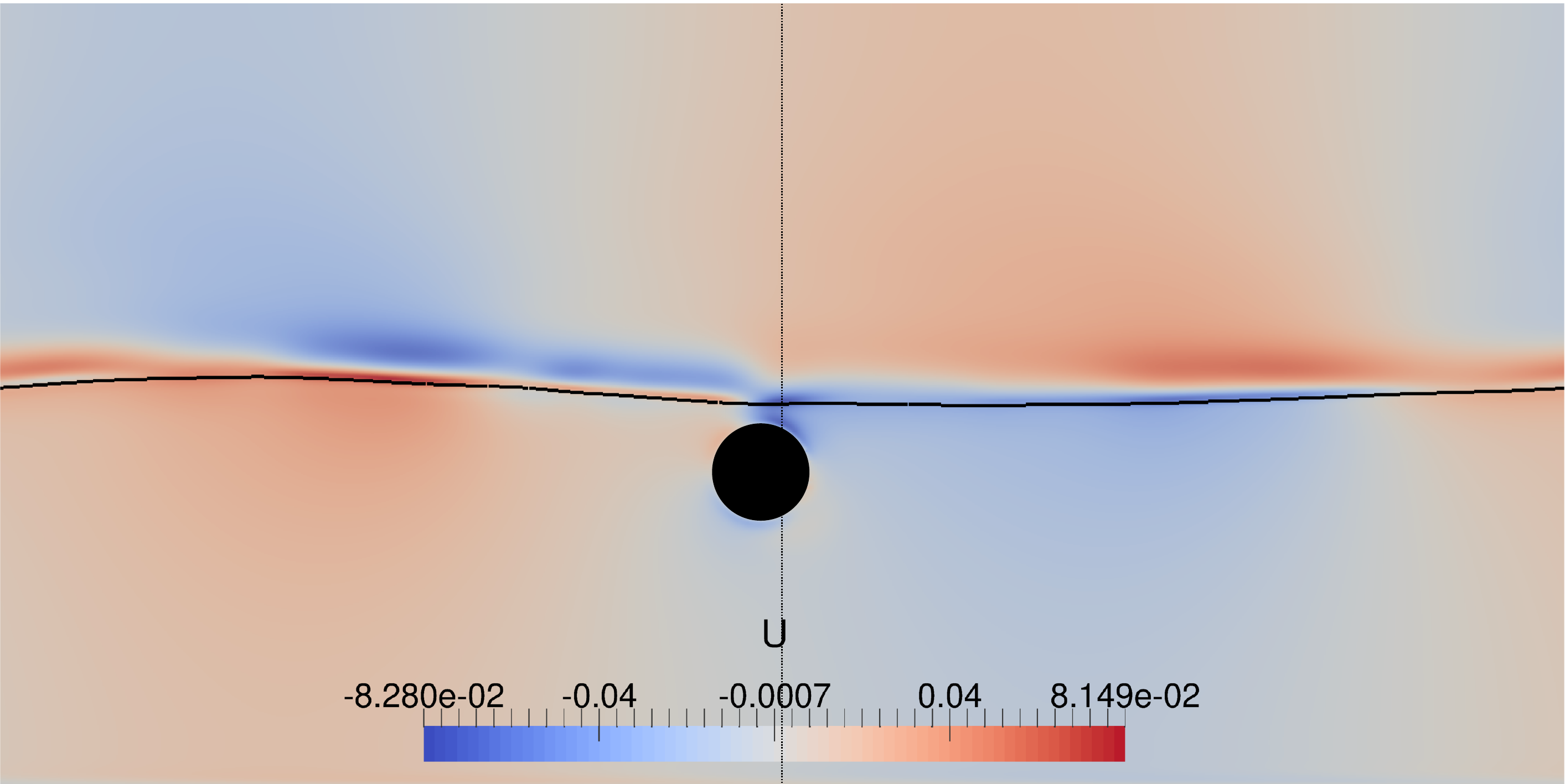}
    \caption{\label{fig:t3}}
  \end{subfigure}
  \caption{Surface gravity wave propagating over a submerged pendulum. (a) horizontal position of the pendulum; black dots are the instants for other panels. (b)-(d) color map of horizontal velocity, position of the pendulum and of the interface (the line has been made thicker in the sake of clarity), corresponding to the black dots in panel (a) with time increasing from top left to bottom right. The entire domain is represented. \label{fig:results}}
\end{figure}

The wave induces an oscillatory motion of the pendulum, as shown in figure 
\ref{fig:pendpos}. Since the wave period is different from that of the pendulum,
the dynamics is characterised by  different oscillation frequencies. Figures 
\ref{fig:t1}-\ref{fig:t2}-\ref{fig:t3} exhibit the contour of the horizontal 
velocity, the interface location and the pendulum position at four time instants, 
corresponding to the black dots in figure \ref{fig:pendpos}. Based on the location 
of the pendulum with respect to the phase of the velocity field, pendulum and wave 
can be in phase or out of phase. The small peaks in the position, for example,
correspond to instants in which the pendulum has a negative velocity and interacts
with the positive phase velocity of the wave, as shown in figure \ref{fig:t2}.

\begin{figure}
  \centering
  \begin{subfigure}{0.49\textwidth}
    \includegraphics[width=\textwidth]{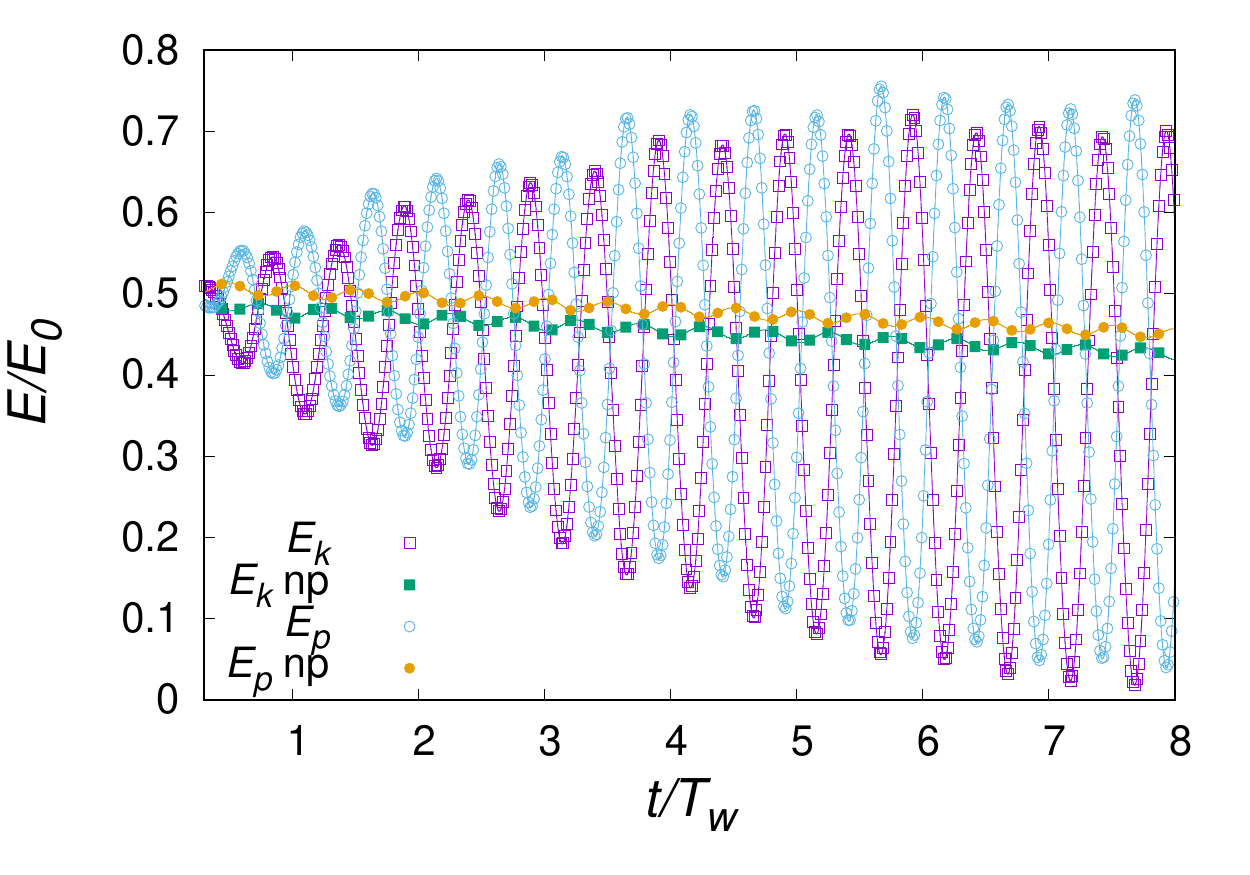}
    \caption{\label{fig:ekp}}
  \end{subfigure}
  \begin{subfigure}{0.49\textwidth}
    \includegraphics[width=\textwidth]{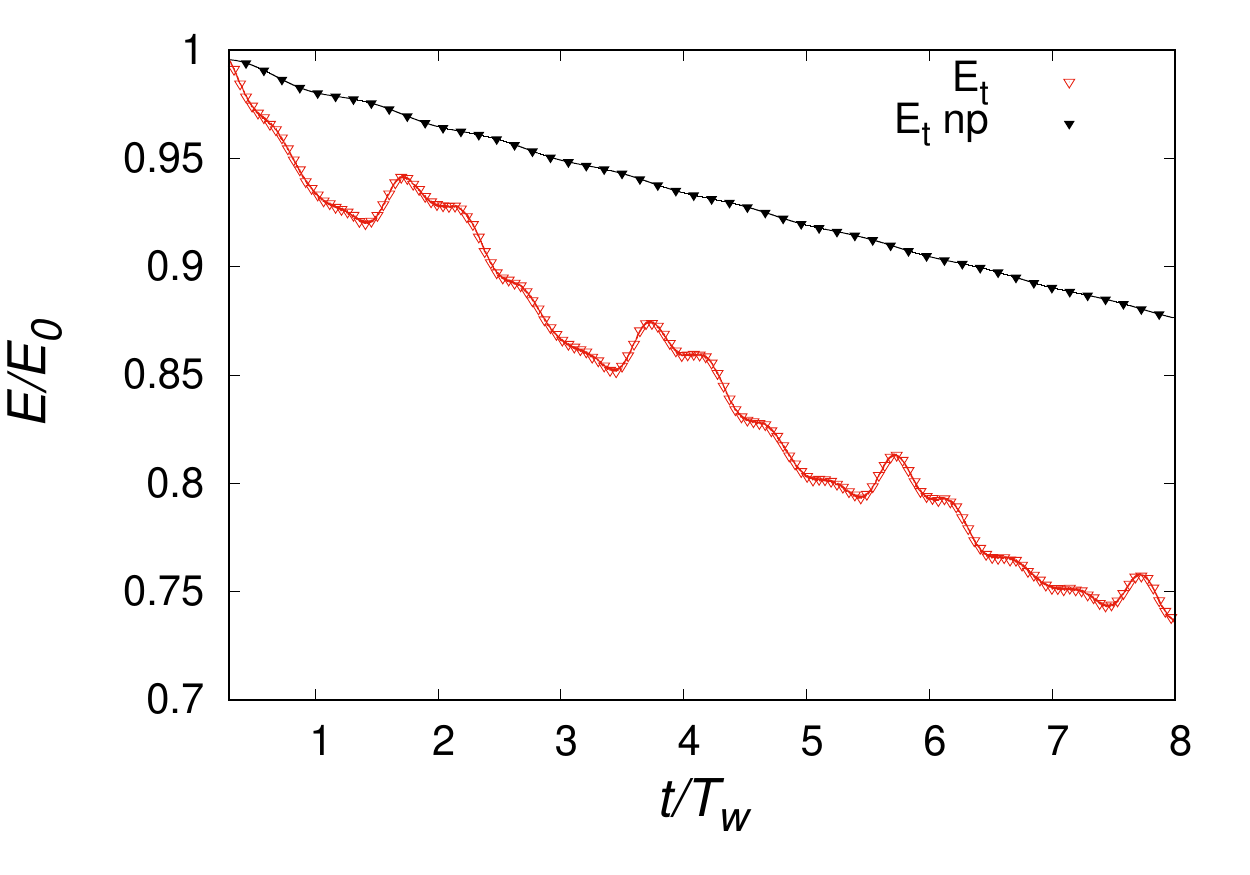}
    \caption{\label{fig:et}}
  \end{subfigure}
  \begin{subfigure}{0.49\textwidth}
    \includegraphics[width=\textwidth]{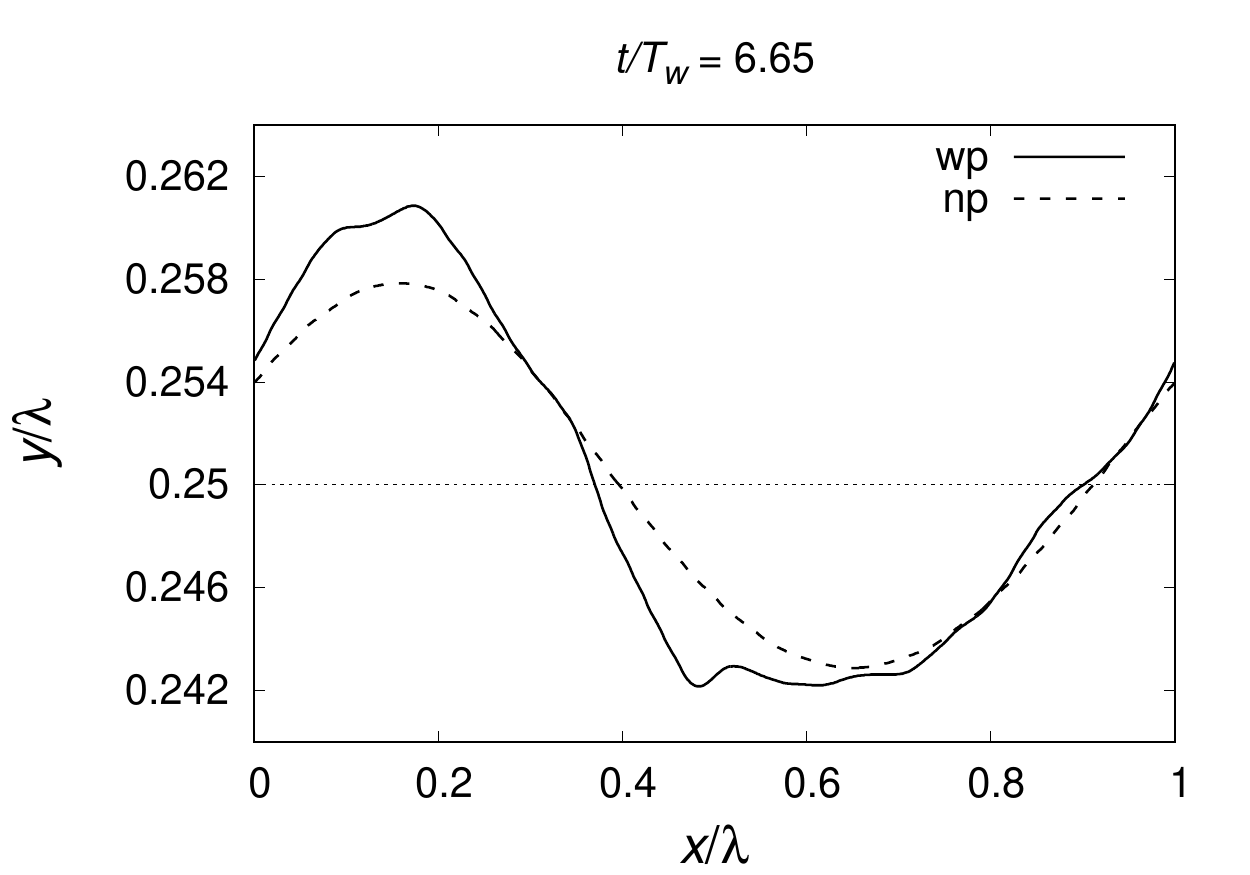}
    \caption{\label{fig:s1}}
  \end{subfigure}
  \begin{subfigure}{0.49\textwidth}
    \includegraphics[width=\textwidth]{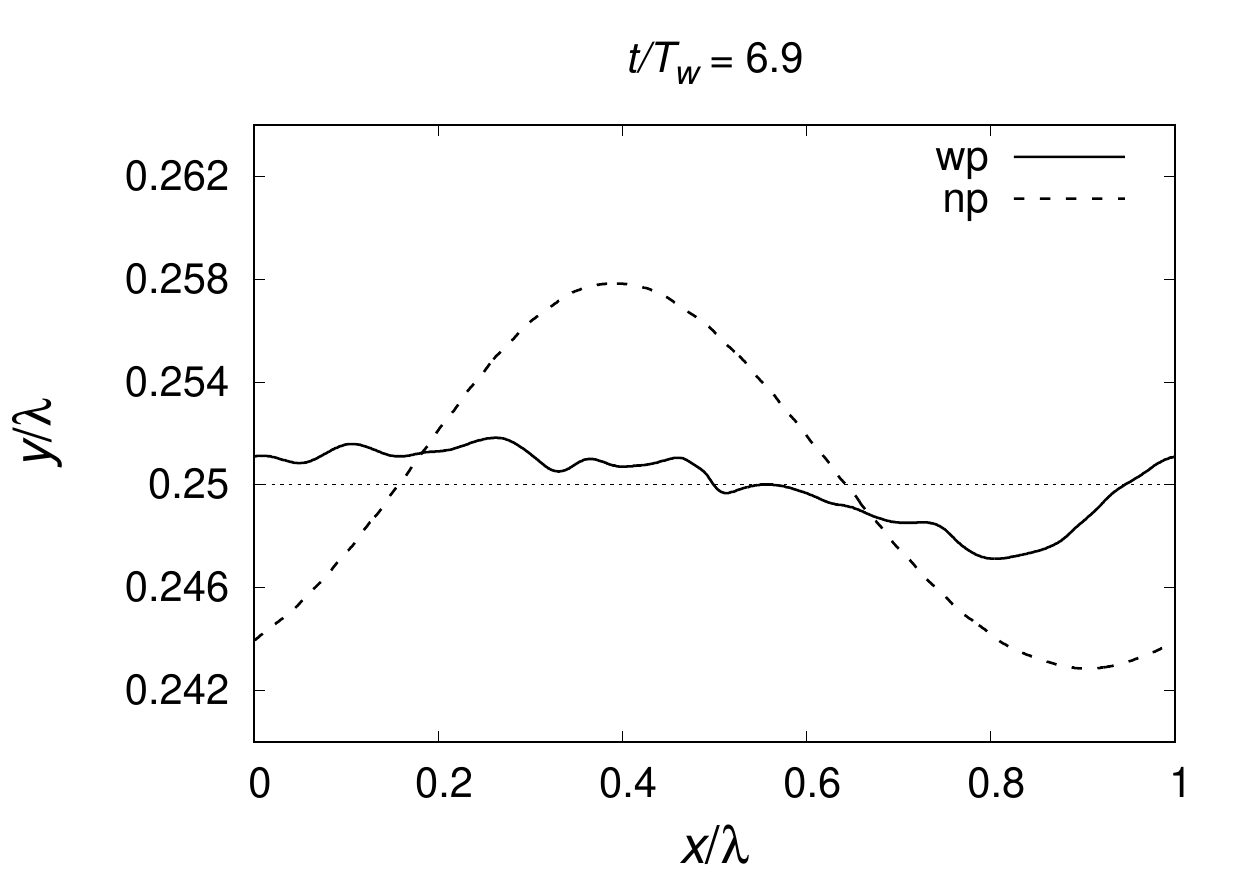}
    \caption{\label{fig:s2}}
  \end{subfigure}
  \caption{Surface gravity wave propagating over a submerged pendulum. 
(a) time history of wave energy components for the case with and without pendulum: 
wave kinetic energy with pendulum (purple open square), 
wave potential energy with pendulum (blue open circles), 
wave kinetic energy without pendulum (green solid square), 
wave potential energy without pendulum (orange solid circles). 
(b) Total wave energy for the case with pendulum (black solid triangles) 
and without pendulum (red open triangles). 
All terms are normalized by the initial total energy. (c)--(d) comparison of 
the surface profile between the case with pendulum (solid line, labeled as wp) 
and without (dashed line, labeled np). 
Dotted line represents the still water level.\label{fig:results2}}
\end{figure}

One interesting aspect of the problem is the energy transfer between the wave and 
the pendulum. We report in figure \ref{fig:ekp} the time history of the kinetic and 
potential energy, computed by equations \eqref{eqn:kin} and \eqref{eqn:pot}, for the 
cases with and without pendulum. For this case the wave potential energy at rest is 
computed only in the fluid domain, \emph{i.e.} 
$E_{p0} = \rho_wg(-\lambda h^2/2+d\pi r^2)$, with $d$ varying in time. 
The presence of the pendulum induces strong oscillations in the wave energy components,
both the kinetic and the potential, which implies a continuous energy transfer 
between the wave and the pendulum. When the wave is out of phase with the pendulum, 
the solid opposes to the wave and induces an increase in the wave height 
(as in figure \ref{fig:s1}) which corresponds to an increase in the potential energy 
of the wave. This energy then goes into kinetic energy with a strong reduction of the 
potential contribution, which in some instant drops almost to zero, corresponding to 
a small wave amplitude. The energy is then dissipated in the fluid by vorticity 
production and the system has an overall higher energy dissipation in time, as 
clearly shown in figure \ref{fig:et}.    

\subsection{3D wave breaking induced by a submerged body}

\begin{figure}
  \centering
  \begin{subfigure}{0.49\textwidth}
    \includegraphics[width=\textwidth]{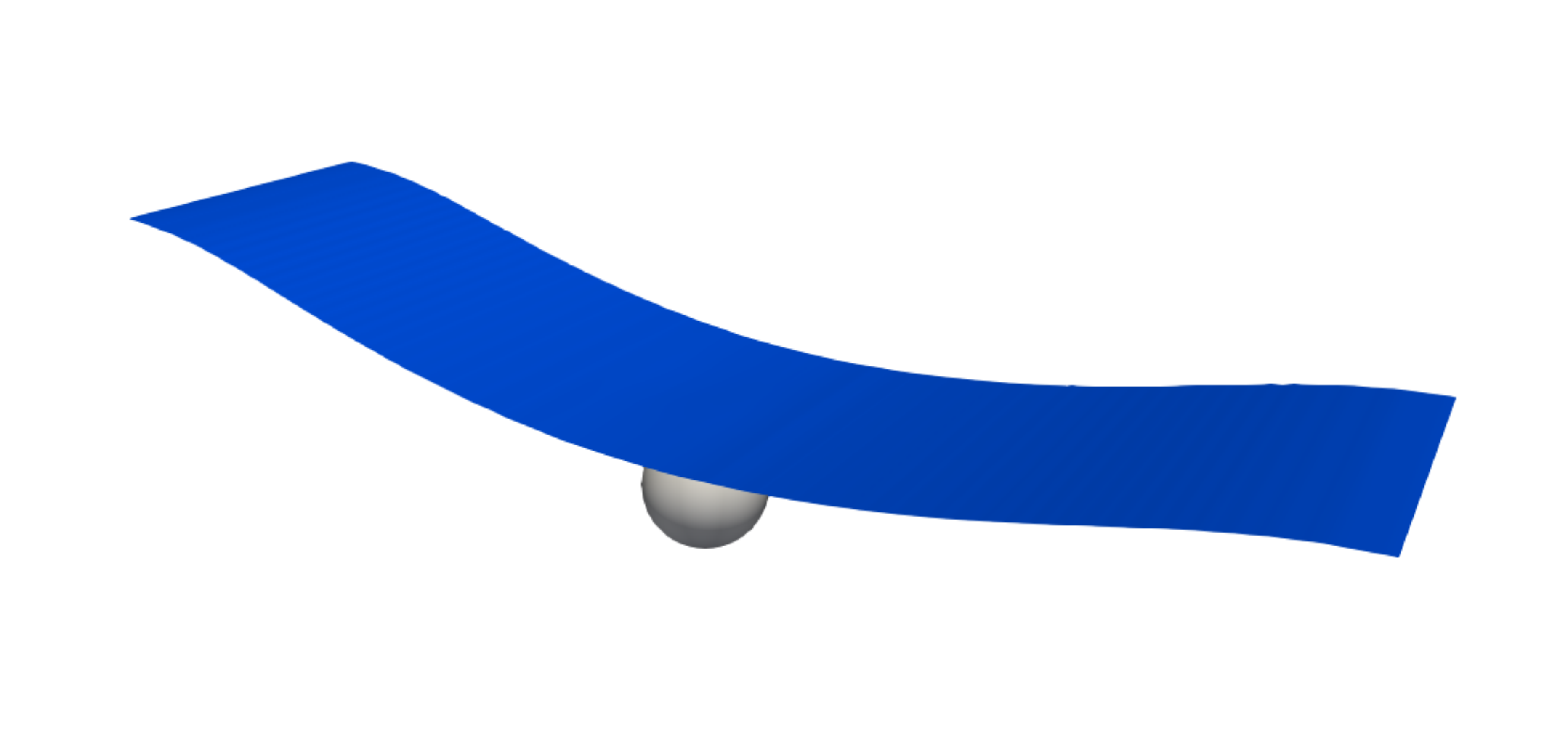}
    \caption{}
  \end{subfigure}
  \begin{subfigure}{0.49\textwidth}
    \includegraphics[width=\textwidth]{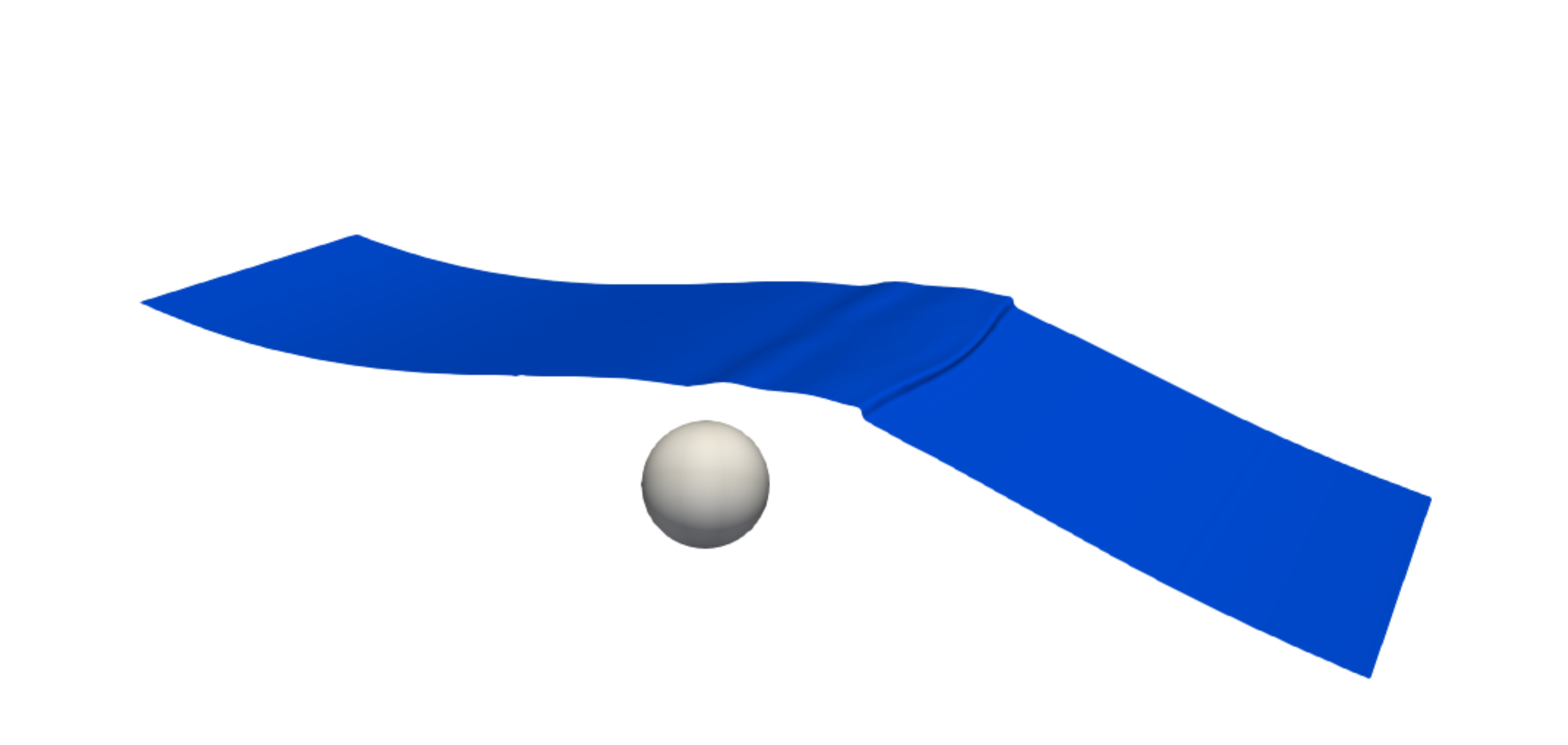}
    \caption{}
  \end{subfigure}
  \begin{subfigure}{0.49\textwidth}
    \includegraphics[width=\textwidth]{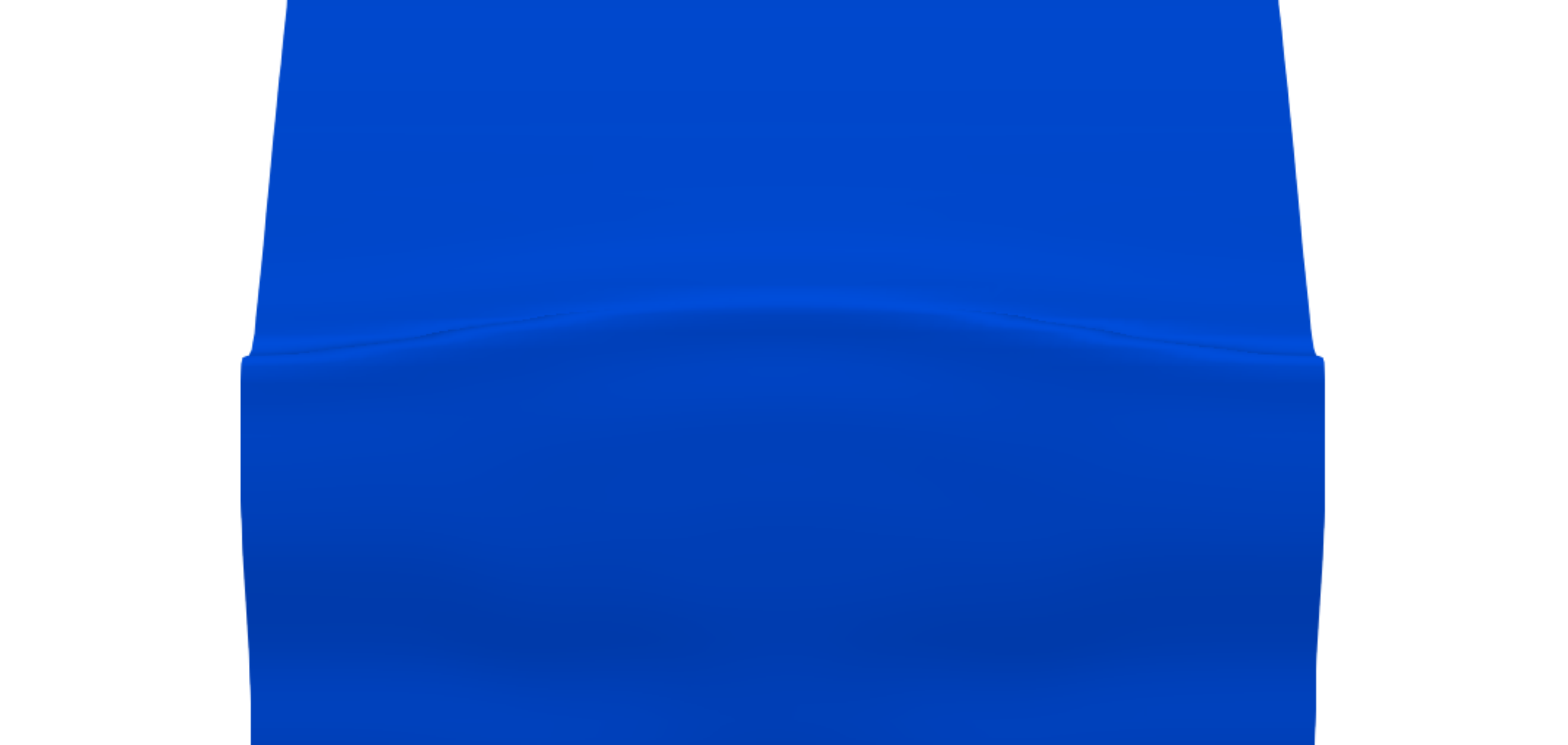}
    \caption{}
  \end{subfigure}
  \begin{subfigure}{0.49\textwidth}
    \includegraphics[width=\textwidth]{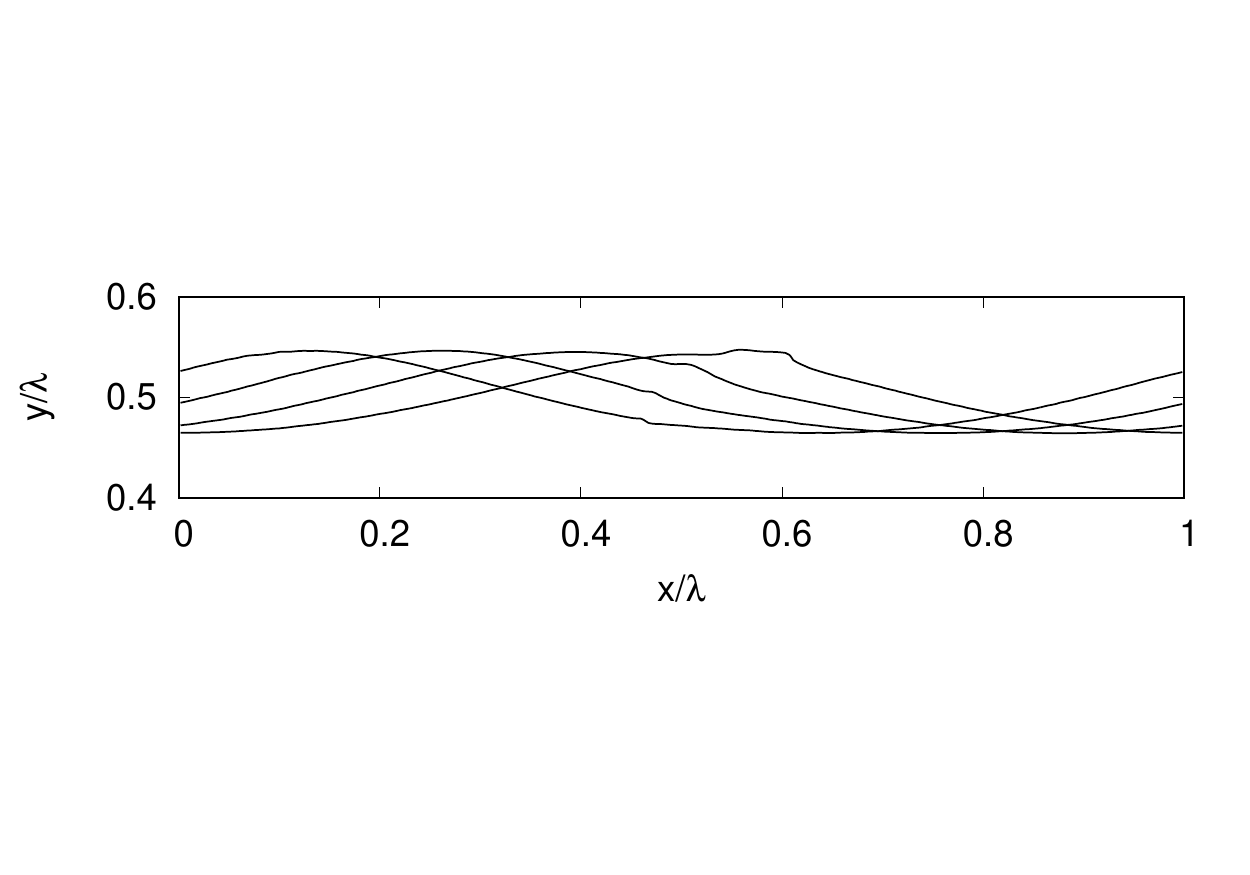}
    \caption{}
  \end{subfigure}
  \caption{3D surface wave breaking: interface location at $T = 0$ (a) and at $T = 0.5$ (b); the wave propagates from the left to the right. Detailed view of the breagkin front from top (c). Wave profile in a $x-y$ plane located at the center in the $z$ direction (d).\label{fig:3D}}
\end{figure}

Finally we want to show a three--dimensional application of the solver. Wave 
breaking is an important phenomenon which occurs in ocean and heavily affects
the air--water mass and energy exchanges. It has been extensively studied in the 
last decades and several criteria have been proposed in order to predict the onset 
of wave breaking \cite{Perlin2013}. Here we want to show the effect of a submerged
body on the breaking of a surface gravity wave. To this aim we simulate a 
third--order Stokes wave 
\begin{equation}
  \eta(x) = \frac{a}{\lambda}\left(\cos(kx) + \frac{1}{2}\varepsilon \cos(2kx) + \frac{3}{8}\varepsilon^2 \cos(3kx)\right),
\end{equation}
which, for a high enough value of the initial steepness 
$\varepsilon = 2\pi a/\lambda$ (about 0.32), leadto wave breaking 
\cite{Chen1999,Iafrati2009,Deike2015}. The submerged body is a sphere of radius 
$r = 0.05 \lambda$ located at a distance $d = 0.1$ below the still water level; 
the initial velocity field is derived from the linear theory and the Reynolds number,
based on the wavelength $\lambda$ and phase speed $c = \sqrt{g\lambda}$,
is set to $Re = \rho \lambda \sqrt{g \lambda}/\mu = 40000$. In this simulation 
the initial steepness is $\varepsilon =0.25$ which corresponds to a non--breaking 
case \cite{Iafrati2009,Deike2015}. However, the presence of the sphere produces a 
breaking front on the surface which is a spilling breaking, as found for larger 
values of the steepness. The sphere produce a disturbance on the surface 
and when the wave crest reaches the location of the sphere (figure \ref{fig:3D} d), it induces an increase in the local height which corresponds to a steepness rise. If this 
perturbation is strong enough to overcome the limiting steepness, wave breaking 
occurs, as clearly shown in figure \ref{fig:3D}(b). The simulation has been 
performed in a domain of size $\lambda \times \lambda \times \lambda/4$ with
a resolution $256$ nodes per wavelength and a computational time of 1 hour per wave period on 512 
processors Intel Xeon E5-2697 v4 at 2.30 GHz.

\section{Conclusions}

A fully Eulerian solver for the solution of multiphase flows with solid bodies is presented which allows an efficient parallelization of the solver. The solid phase is described by an immersed body formulation with a direct forcing approach and an interpolated procedure for the reconstruction of the velocity field close to the solid boundary. The rigid body motion is coupled with the fluid solver through a 4th order predictor corrector method which allows for stable simulations in presence of large oscillations or solids lighter than the surrounding fluid, for which the added mass play an important role. In this case also underelxation could be used to stabilize the simulation. The hydrodynamic loads are computed by means of probes located in the fluid domain close to the forcing nodes and their accurate computation is crucial for a proper evaluation of the solid body dynamics. The two fluids phase are simulated by a volume of fluid solver coupled with the splitting method for the constant coefficient Poisson solver. The immersed boundary method do not pose any restriction on the timestep by its coupling with the splitting procedure require a decrease of the timestep for large density ratio beween the fluid and the solid phase while no restriction is found for buoyancy free bodies. The solver is validated with some test cases available in literature as migration of particle in pressure--driven flow, water exit of a cylidner and the frequency oscillation of a sumberged reversed pendulum. Excellent agreement has been found for all test cases. The method is also applied to a case of a surface gravity wave propagating over a submerged pendulum and the 3D spilling breaking induced by a submerged sphere. The method can be applied to solid body of arbitraty shape by replacing the distance function for the solid geometry description with a ray tracing algorithm. Extension to deformable bodies could be done by replacing the direct forcing apporach with a moving least square method and wil be the subject of future work as for the contact angle between the solid and the liquid. 


\section*{Acknowledgments}
M. Onorato, F. De Lillo and F. De Vita have been funded by Progetto di Ricerca
d'Ateneo CSTO160004, by the ``Departments of Excellence 2018/2022''
Grant awarded by the Italian Ministry of Education, University and
Research (MIUR) (L.232/2016).
M. Onorato and F. De Vita acknowledge also  from EU, 
H2020 FET Open BOHEME grant No. 863179.
The authors also acknowledge CINECA for the computational resources under the 
grant IscraC SGWA.

\section*{Appendix A: Hamming's $4^{th}$ order modified predictor-corrector}\label{sec:A}

At each timestep, once the forces are known, equations \eqref{eqn:veltra}-\eqref{eqn:posrot} are advanced in time using a Hamming's 4th order modified predictor-corrector method \cite{Hamming1959}. Defining $\vec{\mathcal{X}} = \left[\vec{X}, \vec{\Theta}\right]^T$ the unknown vector for the position and $\vec{\mathcal{V}} = \left[\vec{V}, \vec{\Omega}\right]^T$ the unknown vector for the velocity, the evolution is computed by the following steps
\begin{enumerate}
\item \emph{Predictor step}: evaluate the predicted solution $\left(\mathcal{X}^{l+1}_p,\mathcal{V}^{l+1}_p\right)$ at timestep $n+1$
  \begin{itemize}
  \item $\vec{\mathcal{A}}^{l} = \vec{F}^l/m$
  \item $\vec{\mathcal{V}}^{l+1}_p = \vec{\mathcal{V}}^{l-3} + \frac{4}{3}\Delta t\left(2\vec{\mathcal{A}}^{l}-\vec{\mathcal{A}}^{l-1}+2\vec{\mathcal{A}}^{l-2}\right)$
  \item $\vec{\mathcal{X}}^{l+1}_p = \vec{\mathcal{X}}^{l-3} + \frac{4}{3}\Delta t\left(2\vec{\mathcal{V}}^{l}-\vec{\mathcal{V}}^{l-1}+2\vec{\mathcal{V}}^{l-2}\right)$
  \end{itemize}
  improve it with an estimation of the error at previous timestep
  \begin{itemize}
  \item $\vec{\mathcal{V}}^{l+1}_m = \vec{\mathcal{V}}^{l+1}_p - \frac{112}{121}\left(\vec{\mathcal{V}}^{l}_p-\vec{\mathcal{V}}^{l}\right)$
  \item $\vec{\mathcal{X}}^{l+1}_m = \vec{\mathcal{X}}^{l+1}_p - \frac{112}{121}\left(\vec{\mathcal{X}}^{l}_p-\vec{\mathcal{X}}^{l}\right)$
  \end{itemize}
  and solve flow equations using as boundary condition $\left(\mathcal{X}^{l+1}_m,\mathcal{V}^{l+1}_m\right)$ to evaluate $F^{l+1}_1$.
\item \emph{Corrector step}: correct the solution iteratively (loop on $k$, starting from $k = 1$)
  \begin{itemize}
  \item $\vec{\mathcal{A}}^{l+1}_k = \vec{F}^{l+1}_k/m$
  \item $\vec{\mathcal{V}}^{l+1}_k = \frac{1}{8}\left(9\vec{\mathcal{V}}^{l}-\vec{\mathcal{V}}^{l-2}\right) +\frac{3}{8}\Delta t\left(2\vec{\mathcal{A}}^{l+1}_k+2\vec{\mathcal{A}}^{l}-\vec{\mathcal{A}}^{l-1}\right)$
  \item $\vec{\mathcal{X}}^{l+1}_k = \frac{1}{8}\left(9\vec{\mathcal{X}}^{l}-\vec{\mathcal{X}}^{l-2}\right) +\frac{3}{8}\Delta t\left(2\vec{\mathcal{V}}^{l+1}_k+2\vec{\mathcal{V}}^{l}-\vec{\mathcal{V}}^{l-1}\right)$
  \end{itemize}
  and check for convergence: $\min\left(\vec{|\mathcal{X}}^{l+1}_{k}-\vec{\mathcal{X}}^{l+1}_{k-1}|,|\vec{\mathcal{V}}^{l+1}_{k}-\vec{\mathcal{V}}^{l+1}_{k-1}|\right) < \epsilon$, where the values at $k=0$ are the modified value of the predictor step and $\epsilon$ is the tolerance. If converged, make the final correction to the solution:
  \begin{itemize}
  \item $\vec{\mathcal{V}}^{l+1} = \vec{\mathcal{V}}^{l+1}_k + \frac{9}{121}\left(\vec{\mathcal{V}}^{l+1}_p-\vec{\mathcal{V}}^{l+1}_k\right)$
  \item $\vec{\mathcal{X}}^{l+1} = \vec{\mathcal{X}}^{l+1}_k + \frac{9}{121}\left(\vec{\mathcal{X}}^{l+1}_p-\vec{\mathcal{X}}^{l+1}_k\right)$
  \end{itemize}
  and solve flow equations with boundary condition $\left(\mathcal{X}^{l+1},\mathcal{V}^{l+1}\right)$. If not converged, solve flow equations with boundary condition $\left(\mathcal{X}^{l+1}_k,\mathcal{V}^{l+1}_k\right)$, evaluate $F^{l+1}_k$ and then repeat the corrector step until convergence.
\end{enumerate}


\bibliographystyle{authordate1}
\bibliography{biblio}

\end{document}